\documentclass[twocolumn,twocolappendix]{aastex631}

\usepackage{graphicx}	% Including figure files
\usepackage{amsmath}	% Advanced maths commands
\usepackage{multirow}
\usepackage{comment}
\usepackage{array}
\usepackage{soul}
\usepackage{etoolbox}
\usepackage{ae,aecompl}
\usepackage{hyperref}
\usepackage{threeparttable}
\usepackage{mathrsfs}
\usepackage{scalerel}
\usepackage{longtable,booktabs,threeparttablex}
\usepackage{float}

\definecolor{purple}{RGB}{160,32,240}
\definecolor{red}{RGB}{225,50,50}

\definecolor{addchange}{RGB}{215,25,25}

\definecolor{removechange}{RGB}{25,25,215}

\newcommand{\HST}{\emph{HST}}
\newcommand{\JWST}{\emph{JWST}}
\newcommand{\Spitzer}{\emph{Spitzer}}
\newcommand{\Muv}{\ensuremath{M_\mathrm{UV}^{ }}}

\newcommand{\zphot}{\ensuremath{z_{\mathrm{phot}}}}
\newcommand{\zspec}{\ensuremath{z_{\mathrm{spec}}}}

\newcommand{\fesc}{\ensuremath{f_{\mathrm{esc}}}}

\newcommand{\OIIIHb}{[O\,{\sc iii}]+H\ensuremath{\beta}}
\newcommand{\OIII}{[O\,{\sc iii}]}
\newcommand{\OII}{[O\,{\sc ii}]}
\newcommand{\NeIII}{[Ne\,{\sc iii}]}
\newcommand{\NII}{[N\,{\sc ii}]}

\newcommand{\BBrat}{\ensuremath{F_\nu}(4200\,\AA{})\,/\,\ensuremath{F_\nu}(3500\,\AA{})}
\newcommand{\BBshort}{\ensuremath{F_{4200}/F_{3500}}}
\newcommand{\Luv}{\ensuremath{L_\mathrm{UV}}}
\newcommand{\Lrat}{\ensuremath{L_\mathrm{H\alpha}/L_\mathrm{UV}}}
\newcommand{\logLrat}{\ensuremath{\log\left(L_\mathrm{H\alpha}/L_\mathrm{UV}\right)}}

\newcommand{\SFRrat}{\ensuremath{\mathrm{SFR_{3\,Myr}}/\mathrm{SFR_{3-50\,Myr}}}}

\newcommand{\Msol}{\ensuremath{M_{\odot}}}
\newcommand{\Mstar}{\ensuremath{M_{\ast}}}

\newcolumntype{P}[1]{>{\centering\arraybackslash}p{#1}}
\newcommand\Tstrut{\rule{0pt}{2.6ex}}         % = `top' strut
\newcommand\Bstrut{\rule[-1.2ex]{0pt}{0pt}}   % = `bottom' strut

\shorttitle{The Burstiness of Star Formation at $z\sim6$}
\shortauthors{Endsley et al.}

\defcitealias{Endsley2024}{E24}

\begin{document}

\title{The Burstiness of Star Formation at $z\sim6$: \\ A Huge Diversity in the Recent Star Formation Histories of Very UV-faint Galaxies}

\author[0000-0003-4564-2771]{Ryan Endsley}
\affiliation{Department of Astronomy, University of Texas at Austin, 2515 Speedway, Austin, Texas 78712, USA}
\correspondingauthor{Ryan Endsley}
\email{ryan.endsley@austin.utexas.edu}
\author[0000-0002-0302-2577]{John Chisholm}
\affiliation{Department of Astronomy, University of Texas at Austin, 2515 Speedway, Austin, Texas 78712, USA}
\author{Daniel P. Stark}
\affiliation{Steward Observatory, University of Arizona, 933 N Cherry Avenue, Tucson, AZ 85721, USA}
\author[0000-0001-8426-1141]{Michael W. Topping}
\affiliation{Steward Observatory, University of Arizona, 933 N Cherry Avenue, Tucson, AZ 85721, USA}
\author[0000-0003-1432-7744]{Lily Whitler}
\affiliation{Steward Observatory, University of Arizona, 933 N Cherry Avenue, Tucson, AZ 85721, USA}

%% Mark off the abstract in the ``abstract'' environment. 
\begin{abstract}

IRAC data have long implied that early ($z\gtrsim6$) galaxies often have very high specific star formation rates (sSFR$\gtrsim$30 Gyr$^{-1}$), but JWST data have shown that at least some early galaxies are forming stars far less vigorously. Here, we systematically analyze the recent star formation histories (SFHs) of a large ($N=368$) sample of $z\sim6$ Lyman-break galaxies (LBGs) spanning $-22\lesssim M_\mathrm{UV}\lesssim-16$ assembled from ACS+NIRCam imaging in the GOODS and Abell 2744 fields. We find that very low H$\alpha$-to-UV luminosity ratios ($L_\mathrm{H\alpha}/L_\mathrm{UV}$) and strong recent downturns in star formation rate (SFR) are $\approx$5$\times$ more common among the UV-faintest subset of our sample ($\langle M_\mathrm{UV}\rangle=-17.4$) compared to the brightest subset ($\langle M_\mathrm{UV}\rangle=-20.0$). The frequency of high $L_\mathrm{H\alpha}/L_\mathrm{UV}$ and strong recent SFR upturns is approximately constant with UV luminosity. We discuss how bursty SFHs naturally reproduce this much greater diversity in recent SFHs among very UV-faint galaxies. Using public NIRSpec/prism data, we newly confirm recent strong SFR downturns among three LBGs in our sample, and validate our photometric inferences on key SFH signatures among $z\sim6$ LBGs in general. Our results imply that early galaxies frequently cycle through phases of rapid stellar mass assembly and other periods of much slower growth. This yields huge ($\gtrsim$1--2 mag) fluctuations in $M_\mathrm{UV}$ on rapid ($\sim$10--30 Myr) timescales, helping explain the surprising abundance of $z>10$ galaxies. Finally, we caution that this burstiness causes all existing high-redshift samples (particularly line-selected samples) to be far less complete to galaxies with long recent phases of low sSFR than those currently undergoing a burst.

\end{abstract}

%% Keywords should appear after the \end{abstract} command. 
%% The AAS Journals now uses Unified Astronomy Thesaurus concepts:
%% https://astrothesaurus.org
%% You will be asked to selected these concepts during the submission process
%% but this old "keyword" functionality is maintained in case authors want
%% to include these concepts in their preprints.
\keywords{High-redshift galaxies(734) --- Galaxy formation(595) --- Galaxy evolution(594)}

\section{Introduction} \label{sec:intro}

The star formation histories (SFHs) of galaxies within the first billion years of cosmic history ($z\gtrsim6$) encode the physics that regulated galaxy formation.
For much of the past two decades, our understanding of star formation at $z\gtrsim6$ was almost solely restricted to the UV-bright ($\Muv{} \lesssim -20$) galaxy population due to the severe sensitivity limitations of \Spitzer{}/IRAC.
These IRAC-based studies repeatedly found signatures of high equivalent width (EW) rest-optical line emission among $z\gtrsim6$ galaxies \citep[rest-frame EW$\gtrsim$700 \AA{}; e.g.,][]{Labbe2013,Smit2014,Castellano2017,deBarros2019,Endsley2021_OIII,Stefanon2022_z8Ha,Stefanon2022_sSFR}, far more frequently than seen at even $z\sim2$ \citep{Boyett2022_OIII}.
Such findings implied a strong increase in galaxy specific star formation rates (sSFRs) towards high redshifts, broadly consistent with model predictions of halo growth \citep[e.g.,][]{Dekel2014,Tacchella2018}.

However, it was unclear whether the ubiquitous high-EW emission lines seen at $z\gtrsim6$ were due to steady efficient halo growth, or instead due to `bursty' star formation histories.
In bursty SFHs, the star formation rate (SFR) of a galaxy fluctuates strongly over short time intervals.
Consequently, a substantial fraction of the total stellar mass is formed during recurrent, short-lived spikes in SFR (i.e., bursts).
These burst episodes temporarily make a galaxy much brighter while also producing higher EW line emission due to the sudden formation of many O stars, which have enormous light-to-mass ratios and ionizing photon production efficiencies.

Prior to the launch of \JWST{}, several simulations and theoretical models predicted that early galaxies commonly experienced bursty SFHs (e.g., \citealt{Dayal2013,Kimm2015,Ceverino2018,FaucherGiguere2018,Ma2018_burstySFH,Rosdahl2018,Barrow2020,Furlanetto2022}).
At the time, IRAC data had also revealed that a substantial fraction ($\approx$20\%) of UV-bright $z\sim7-8$ galaxies exhibited extremely high \OIIIHb{} EWs ($>$1200 \AA{} rest-frame;  \citealt{Smit2014,RobertsBorsani2016,Endsley2021_OIII}).
In the context of star formation, such powerful \OIIIHb{} nebular line emission relative to the underlying continuum indicates that nearly all of the UV and optical light is powered by early-type O stars.
These extremely high EWs thus implied rapid, strong upturns in SFR within the past $\sim$3 Myr \citep[e.g.,][]{Whitler2023_z7} consistent with an ongoing `burst' event.

Nonetheless, IRAC's limited capabilities completely prohibited detailed SFH characterization at $z\gtrsim6$.
Two crucial shortcomings prohibited IRAC data from revealing if early galaxies had recently experienced a strong downturn in SFR, as would be expected shortly after a burst episode.
First, any galaxy with a strong recent downturn in SFR would exhibit low optical-line EWs (due to the dearth of early-type O stars), yielding flat broadband IRAC colors that could often be matched by models with constant SFR for the past $\gtrsim$100 Myr (see \citealt{Endsley2023_CEERS}).
Second, IRAC simply did not have the sensitivity to readily detect $z\gtrsim6$ galaxies with strong recent SFR downturns.
Such galaxies would necessarily be relatively faint in the continuum since they lack the most massive stars with the highest light-to-mass ratios \citep{Sun2023_observability}.
Overall, the UV-bright and high-EW $z\gtrsim6$ galaxy population revealed by IRAC only offered a very narrow glimpse on early SFHs, making it impossible to quantify the amplitude and cadence of bursts in the first billion years.

The revolutionary capabilities of \JWST{} have enabled rapid progress in empirically constraining $z\gtrsim6$ SFHs, with much attention focused on the previously unseen population of galaxies caught shortly after a burst.
\citet{Endsley2023_CEERS} identified the first candidate $z\sim7-8$ galaxies with strong recent downturns in SFR\footnote{There are several emerging terms for these types of early galaxies, including `mini-quenched' \citep{Dome2024,Looser2024}, `lulling' \citep{Looser2023b}, `napping' \citep{Cole2023}, and `smouldering' \citep{Trussler2024}. To be clear in how we interpret the physical nature of these systems, we simply refer to them as galaxies with a recent strong downturn (or decline) in SFR.} using the greatly improved sensitivity and spectral energy distribution (SED) sampling of the Near-Infrared Camera (NIRCam; \citealt{Rieke2005_NIRCam,Rieke2023_NIRCam}).
These galaxies showed 1) weak/non-existent Balmer breaks indicating continua powered by OB stars, and 2) surprisingly weak \OIIIHb{} lines implying relatively little star formation over the past $\sim$3 Myr.
Similar candidates were later presented in \citet[][hereafter \citetalias{Endsley2024}]{Endsley2024} at $z\sim6$.
The weak H$\alpha$ emission seen at this slightly lower redshift indicated that the low \OIIIHb{} EWs were not simply due to very metal-poor gas weakening \OIII{}.
Recently, \citet{Looser2024} demonstrated that early galaxies can experience very low sSFRs over a considerably longer timescale ($\sim$20 Myr).
Using extremely deep Near-Infrared Spectrograph (NIRSpec; \citealt{Jakobsen2022_NIRSpec,Boker2023_NIRSpec}) data, they report a $z=7.3$ galaxy with a clear Balmer break, no \OIII{} emission, and Balmer-line absorption indicating an effectively complete absence of O stars, and perhaps even few early-type B stars.

Over the past two years, there have been several other \JWST{}-based papers aiming to constrain the SFHs of $z\gtrsim6$ galaxies (e.g., \citealt{Cole2023,Dressler2023,Dressler2024,Looser2023b,Tacchella2023_NIRCamNIRSpec,Whitler2023_z10,Asada2024,Caputi2024,Clarke2024,Ciesla2024,GimenezArteaga2024,Harshan2024,Trussler2024,Wang2024_RUBIESb,Weibel2024}).
While nearly all of these works agree that early galaxies show indications of `bursty' SFHs, considerable effort remains to reach a clear consensus on the amplitude and cadence of these bursts.
An ultimate goal is to use empirical SFH constraints to test the physics implemented in various models of galaxy formation.
The `burstiness' predicted by a given model is highly sensitive to the timing of feedback after the onset of star formation, and the efficiency of feedback in driving gas out of the interstellar medium \citep[e.g.,][]{Hopkins2014,Furlanetto2022,Bhagwat2024,Pallottini2024}.
The SFHs of high-redshift galaxies are therefore intricately tied to our understanding of early galactic ecosystems.

The overarching goal of this paper is to improve our understanding of the `burstiness' of early star formation.
To do so, we analyze a statistical sample ($N = 368$) of $z\sim6$ Lyman-break galaxies (LBGs) selected using very deep JWST/NIRCam imaging in the Abell 2744 lensing field, as well as the Great Observatories Origins Deep Survey (GOODS) blank fields.
These fields all possess overlapping coverage in several broad and medium bands, enabling rich SED constraints even for systems with relatively low light-to-mass ratios due to a recent downturn in SFR.
From this sample, we systematically identify and characterize LBGs with the most confident recent strong SFR downturns (using public NIRSpec/prism data where available), discussing the inferred lookback time to their most recent burst. 
We then use the full LBG sample to quantify how the H$\alpha$ to far-UV luminosity ratio correlates with UV luminosity.
This luminosity ratio, \Lrat{}, has long been used as a proxy for burstiness among star-forming galaxies \citep[e.g.,][]{Glazebrook1999,daSilva2014,Emami2019,Faisst2019,Atek2022}, relying on the principle that H$\alpha$ emission is powered by O stars formed over the past $\approx$10 Myr, while the far-UV continuum is also powered by B stars that live for $\sim$100 Myr. 
Finally, we infer the distribution of recent SFH shapes (i.e., the fraction of strong recent SFR upturns and SFR downturns) as a function of UV luminosity.

This paper is structured as follows.
We begin by detailing the datasets used in this work, our LBG selection, and the photoionization SED modelling used to infer the recent SFHs (\S\ref{sec:sec2}).
Next, we utilize deep public NIRSpec/prism data to explicitly test (and ultimately validate) our photometric SED inferences (\S\ref{sec:spectra}), including for objects with inferred strong recent SFR downturns (\S\ref{sec:specDownturns}).
Next, we quantify how the H$\alpha$ to far-UV luminosity ratios (\S\ref{sec:Ha_UV}) and recent SFH shapes (\S\ref{sec:SFRratDistnsMuv}) of $z\sim6$ LBGs correlate with UV luminosity.
We discuss the implications of our findings in \S\ref{sec:discussion} and our main conclusions are summarized in \S\ref{sec:summary}.

Throughout this paper, we quote all magnitudes in the AB system \citep{Oke1983}, report EWs in rest-frame units, assume a \citet{Chabrier2003} stellar initial mass function (IMF) with limits of 0.1--300 \Msol{}, and adopt a flat $\Lambda$CDM cosmology with parameters $h=0.7$, $\Omega_\mathrm{M}=0.3$, and $\Omega_\mathrm{\Lambda}=0.7$. 

\begin{figure}
\includegraphics[width=\columnwidth]{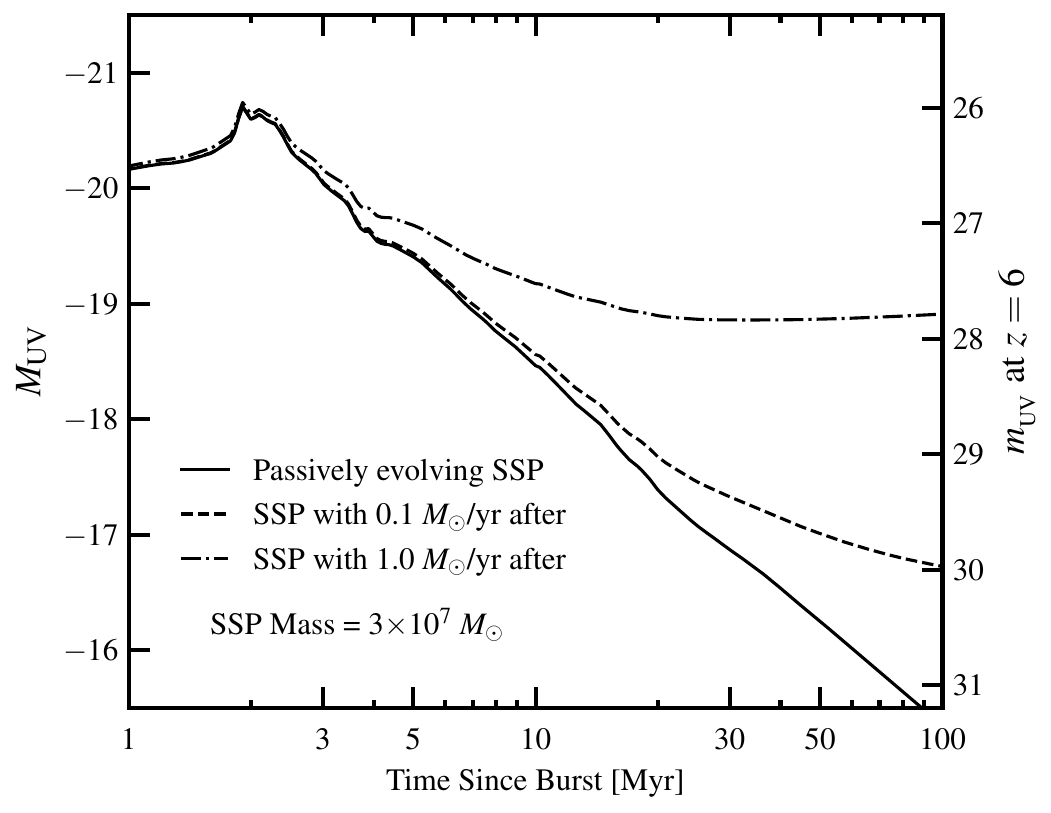}
\caption{Illustration of how rapidly a galaxy's far-UV luminosity (at rest-frame 1500 \AA{}) declines after a burst of star formation. The left ticks show the absolute UV magnitude while the right ticks show the apparent magnitude at $z=6$. The solid line shows the relation for a single passively-evolving simple stellar population (SSP) with a mass of $3\times10^7$ $M_\odot$, while the dashed (dot-dashed) line shows the trend for the same SSP but with residual constant star formation after the burst at a level of 0.1 $M_\odot$/yr (1.0 $M_\odot$/yr). These curves were made using the \citet{Gutkin2016} stellar photoionization models assuming a metallicity of 0.1 $Z_\odot$, an ionization parameter of log $U = -2.5$, and no dust attenuation.}
\label{fig:SSPevolution}
\end{figure}

\section{Observations and Galaxy Sample} \label{sec:sec2}

The main goal of this paper is to better understand the recent SFHs of $z\sim6$ galaxies.
Accordingly, we aim to be as complete as possible to galaxies with strong recent \textit{downturns} in SFR, not just those with strong recent upturns that have long been characterized with IRAC (e.g., \citealt{Smit2014,RobertsBorsani2016,Endsley2021_OIII}).
Achieving high completeness to systems with strong recent SFR downturns is a relatively challenging task simply because of how rapidly the light-to-mass ratio plummets as the O stars die off after a spike in SFR.
This necessitates the use of deep \JWST{} data.

To illustrate this point, Fig. \ref{fig:SSPevolution} shows how the far-UV luminosity (at rest-frame 1500 \AA{}) of a galaxy evolves after a strong burst of star formation.
Here we adopt the stellar photoionization models of \citet{Gutkin2016}, which include nebular continuum emission (see \S\ref{sec:photoionization} for details).
Adopting the binary stellar photoionization models from Binary Population and Spectral Synthesis (\textsc{bpass}; \citealt{Stanway2018}) does not significantly change the trend shown in Fig. \ref{fig:SSPevolution} (see Appendix \ref{app:binaries}).
A $z=6$ galaxy that forms 3$\times$10$^7$ $M_\odot$ of stars in an instantaneous burst event will appear as a UV-bright (apparent AB magnitude $m_{\scaleto{\mathrm{UV}}{4pt}}\approx26.5$) object over the subsequent 3 Myr (with a peak around $\approx$2 Myr due to the post-main sequence evolution of the most massive stars).
If there is very little star formation following that burst, the galaxy will turn into a very UV-faint ($m_{\scaleto{\mathrm{UV}}{4pt}}\approx29$) object in just 10--20 Myr.
Such very faint galaxies can only be well characterized in the deepest existing \JWST{} data, or with the assistance of substantial gravitational magnification. 
For this reason, we restrict our analysis to the GOODS northern (GOODS-N) and southern (GOODS-S) fields \citep{Giavalisco2004}, as well as the Abell 2744 (A2744) lensing field. 

In this section, we begin by describing the imaging and spectroscopic data sets used in this work, as well as our data reduction procedures (\S\ref{sec:observations}).
Next, we describe our selection of $z\sim6$ LBGs from the imaging data (\S\ref{sec:LBGselection}) and then our SED modelling approach to inferring the recent SFHs of these galaxies (\S\ref{sec:photoionization}).
We then summarize the photometrically-inferred SED properties of our final sample of 368 $z\sim6$ LBGs (\S\ref{sec:LBGproperties}) and describe the subset of LBGs that have most confidently experienced a recent strong SFR downturn (\S\ref{sec:strongDownturnLBGs}).

\subsection{Observational Data} \label{sec:observations}

As of February 2024, A2744 has been imaged in every NIRCam broad and medium band thanks to several Cycle 1 and 2 programs: GLASS \citep{Treu2022_glass}, Ultradeep NIRSpec and NIRCam ObserVations before the Epoch of Reionization (UNCOVER; \citealt{Bezanson2022}), MegaScience \citep{Suess2024}, All the Little Things (ALT; PID 3516, PIs J. Matthee \& R. Naidu), Medium-band Astrophysics with the Grism of NIRCam in Frontier Fields (MAGNIF; PID 2883, PI F. Sun), PID 2756 (PIs W. Chen, T. Treu, \& H. Williams), PID 3538 (PI E. Iani), and PID 3990 (PI T. Morishita).
We reduce the NIRCam imaging data in A2744 following the procedures outlined in \citetalias{Endsley2024}, using the \JWST{} Calibration Reference Data System context map {\tt jwst\_1188.pmap}.

Deep \textit{Hubble Space Telescope} (\HST) Advanced Camera for Surveys (ACS) imaging in F435W, F606W, and F814W also exists in A2744 from several programs (see Table 3 of \citealt{Bezanson2022}). 
We use the \textit{Gaia}-aligned ACS mosaics from the Dawn JWST Archive (DJA; \citealt{Valentino2023_DJA}) in this work.
In our analysis of $z\sim6$ LBGs in A2744, we adopt the high-resolution $z=6$ lensing magnification map released by the UNCOVER team (v2.0; \citealt{Furtak2023_A2744,Price2024}).
The 16th, 50th, and 84th percentile magnification factors\footnote{Two LBGs in our sample (A2744-23813 and A2744-23814) are very nearby (0.3\arcsec{} separation) and highly magnified ($\mu > 10$). These two systems are very unlikely to be the same multiply-imaged system, and we adopt their analytic magnification factors throughout this work (L. Furtak priv. comm.; see Table \ref{tab:properties}).} inferred among our $z\sim6$ LBG sample in A2744 (see \S\ref{sec:LBGselection}) are $\mu = 1.6$, 2.0, and 3.8, respectively.

For the GOODS-N and GOODS-S fields, we utilize the public catalog of $z\sim6$ LBGs from \citetalias{Endsley2024}.
Briefly, this photometric catalog was constructed using all \JWST{} Advanced Deep Extragalactic Survey (JADES; \citealt{Eisenstein2023}) NIRCam data prior to Feb. 10, 2023 in F090W, F115W, F150W, F200W, F277W, F335M, F356W, F410M, and F444W (see \citealt{Rieke2023_JADES}), as well as overlapping F444W imaging from the First Reionization Epoch Spectroscopically Complete Observations (FRESCO; \citealt{Oesch2023_FRESCO}) program. 
\citetalias{Endsley2024} used \HST{}/ACS mosaics from the \textit{Hubble} Legacy Field archive (HLF; see \citealt{Illingworth2013,Whitaker2019} and references therein) that were registered to the \textit{Gaia} frame (see \citealt{Williams2023_jems}). 

We perform source extraction and photometric measurements in A2744 following the procedures in \citetalias{Endsley2024}, which were applied to construct the GOODS photometric catalog used in this work.
Briefly, we first convolve all A2744 mosaics to the point spread function (PSF) of the band with the broadest PSF (F480M) and then ran \textsc{Source Extractor} \citep{Bertin1996} on an inverse-variance weighted stack of all PSF-convolved long-wavelength NIRCam bands (F250M through F480M).
Photometry is then computed on the PSF-matched mosaics in elliptical Kron apertures using a two-stage aperture correction with photometric errors computed using randomly-placed apertures in nearby empty regions.
A neighbor-subtraction algorithm \citep{Endsley2023_CEERS} was employed to remove contaminating flux from nearby objects that were not found to lie at $z\sim6$.
For $z\sim6$ galaxies identified as separate objects with \textsc{Source Extractor} but with overlapping Kron apertures, we computed photometry in the smallest elliptical aperture containing all pixels from their combined segmentation maps, after subtracting off flux from other neighbors where appropriate.

In Table \ref{tab:A2744depths}, we report the 10th, 50th, and 90th percentile 
5$\sigma$ photometric depths (before any lensing correction) among our A2744 $z\sim6$ LBG sample (described in \S\ref{sec:LBGselection}).
The typical depths are $m_{5\sigma} \approx 28.5$ in ACS, $m_{5\sigma} \approx 29$ in NIRCam broad-bands, and $m_{5\sigma} \approx 28-28.5$ in NIRCam medium-bands.
Table 1 of \citetalias{Endsley2024} provides the same information for the GOODS $z\sim6$ LBG sample used in this work.
The typical depths in A2744 and GOODS are broadly similar in the ACS bands, but the GOODS imaging is $\approx$0.5--1 mag deeper in most NIRCam bands covered by JADES. 
However, the median lensing magnification factor of $\mu = 2.0$ for our A2744 $z\sim6$ LBG sample effectively results in similar typical NIRCam depths.

\begin{table}
\centering
\caption{Summary of the 5$\sigma$ depths achieved for our final Lyman-break $z\sim6$ galaxy sample in Abell 2744. Because we utilize elliptical Kron apertures with source-dependent sizes, we report the 10th, 50th, and 90th percentile depths (not corrected for lensing) in each band. The percentile depths for the GOODS sample are provided in Table 1 of \citetalias{Endsley2024}.}
\begin{tabular}{P{1.2cm}P{1.8cm}P{1.8cm}P{1.8cm}}
\hline
Band & 10$^\mathrm{th}$ percentile & 50$^\mathrm{th}$ percentile & 90$^\mathrm{th}$ percentile \\[2pt]
\hline
F435W & 27.1 & 28.4 & 29.0 \Tstrut{}\\[1pt]
F606W & 27.7 & 28.6 & 29.1\\[1pt]
F814W & 27.4 & 28.7 & 29.3\\[1pt]
F070W & 27.9 & 28.6 & 29.2\\[1pt]
F090W & 28.4 & 29.0 & 29.6\\[1pt]
F115W & 28.2 & 28.8 & 29.3\\[1pt]
F140M & 27.1 & 27.8 & 28.3\\[1pt]
F150W & 28.1 & 28.8 & 29.5\\[1pt]
F162M & 27.1 & 27.9 & 28.4\\[1pt]
F182M & 27.7 & 28.4 & 29.0\\[1pt]
F200W & 28.2 & 29.0 & 29.5\\[1pt]
F210M & 27.6 & 28.3 & 28.8\\[1pt]
F250M & 27.3 & 27.9 & 28.4\\[1pt]
F277W & 28.5 & 29.1 & 29.7\\[1pt]
F300M & 27.8 & 28.3 & 28.9\\[1pt]
F335M & 27.8 & 28.4 & 28.9\\[1pt]
F356W & 28.7 & 29.3 & 29.8\\[1pt]
F360M & 27.7 & 28.3 & 28.9\\[1pt]
F410M & 27.9 & 28.5 & 28.9\\[1pt]
F430M & 27.0 & 27.5 & 28.0\\[1pt]
F444W & 28.3 & 28.8 & 29.4\\[1pt]
F460M & 26.7 & 27.3 & 27.7\\[1pt]
F480M & 26.5 & 27.1 & 27.6\\[1pt]
\hline
\end{tabular}
\label{tab:A2744depths}
\end{table}

To test our photometric inferences on the recent SFHs of $z\sim6$ LBGs, we utilize NIRSpec/prism datasets in the GOODS and A2744 fields available in the public archive. 
In GOODS, we use NIRSpec data from the JADES \citep{Bunker2023_JADES_DR1,DEugenio2024_JADES_DR3} and JADES Origins Field (JOF; \citealt{Eisenstein2023_JOF}) programs, while we use data from the UNCOVER program \citep{Price2024} in A2744.
For consistency in our analysis between the two fields, we independently reduce the NIRSpec spectra following the procedure described in \citet{Topping2024_RXCJ}, which utilizes a combination of the JWST calibration pipeline \citep{Bushouse2023} and custom scripts to subtract `snowballs' and `showers', correct for 1/$f$ noise \citep{Rauscher2024}, subtract background flux, and stack all data for a given object onto a common world coordinate system and wavelength solution. 
From the resulting 2D spectra and its associated uncertainty array, we extract the 1D spectra and noise using optimal extraction \citep{Horne1986}.
We describe our analysis of this public NIRSpec/prism data in \S\ref{sec:spectra} after first describing our Lyman-break galaxy selection and sample.

\subsection{Selection of \texorpdfstring{$z\sim6$}{z ~ 6} Lyman-break Galaxies} \label{sec:LBGselection}

We select $z\sim6$ galaxies using the Lyman-break method \citep[e.g.,][]{Steidel1999,Bunker2004}.
This method simply relies on the presence of an extremely sharp spectral discontinuity at rest-frame 1216 \AA{} due to strong absorption+scattering from neutral hydrogen in the intervening intergalactic medium \citep[e.g.,][]{Inoue2014}.
We intentionally do not employ a high-redshift probability cut in our selection as doing so would bias the recent SFHs of our galaxy sample at fixed UV luminosity (see \citetalias{Endsley2024}).
As described in \S\ref{sec:photoionization}, we do enforce a photometric redshift ($\zphot{}$) cut to remove objects likely at $z>6.6$ where H$\alpha$ falls outside the NIRCam coverage.

In this work, we use the public $z\sim6$ LBG sample in the GOODS fields from \citetalias{Endsley2024} which consists of 278 galaxies at $\zphot{}\approx5.5-6.5$.
Briefly, this sample was selected to show the following colors:
\begin{itemize}
    \item F775W $-$ F090W $>$ 1.2
    \item F090W $-$ F150W $<$ 1.0
    \item F775W $-$ F090W $>$ F090W $-$ F150W $+$ 1.2
\end{itemize}
where the F775W flux is set to its 1$\sigma$ upper limit in cases where S/N(F775W)$<$1.
Additionally, \citetalias{Endsley2024} enforced S/N$<$2 in F435W, along with F606W $-$ F090W $>$ 2.7 (F606W $-$ F090W $>$ 1.8) if S/N(F606W)$>$2 (S/N(F606W)$<$2).
These cuts in S/N(F435W) and F606W $-$ F090W color were ignored for objects with extremely strong Ly$\alpha$ breaks (F775W $-$ F090W $>$ 2.5).
Every $z\sim6$ LBG in the \citetalias{Endsley2024} sample was required to have a $>$5$\sigma$ detection in at least one NIRCam band.
Additionally, \citetalias{Endsley2024} required $>$3$\sigma$ detections in at least three NIRCam bands, as well as either ACS/F814W or ACS/F850LP.
Every selected $z\sim6$ LBG was visually inspected in all HST and NIRCam mosaics to remove spurious sources due to artifacts or diffuse emission from very extended low-redshift objects (\citetalias{Endsley2024}).
Finally, only galaxies with F335M imaging coverage (sensitive to \OIIIHb{} at $z\sim6$) were included in this sample to ensure robust photometric SED constraints.

Our $z\sim6$ LBG selection in A2744 largely follows the same approach, with the exception that we utilize F070W instead of F775W. 
Because there is a larger wavelength separation between F070W and F090W, we enforce a stricter Lyman-$\alpha$ break color cut to better exclude low-redshift contaminants.
Specifically, we enforce:
\begin{itemize}
    \item F070W $-$ F090W $>$ 1.5
    \item F090W $-$ F150W $<$ 1.0
    \item F070W $-$ F090W $>$ F090W $-$ F150W $+$ 1.5.
\end{itemize}
Unless the object has an extremely strong Ly$\alpha$ break (F070W $-$ F090W $>$ 2.5), we also enforce S/N$<$2 in F435W along with F606W $-$ F090W $>$ 2.7 (F606W $-$ F090W $>$ 1.8) if S/N(F606W)$>$2 (S/N(F606W)$<$2).
Finally, we ensure that each selected $z\sim6$ galaxy in A2744 is observed in every NIRCam broad and medium band, has a $>$5$\sigma$ detection in at least one NIRCam band, and a $>$3$\sigma$ detection in at least three NIRCam bands.
This selection results in 102 $z\sim6$ LBGs in A2744.

\subsection{Photoionization SED Modelling} \label{sec:photoionization}

To infer the properties (e.g., recent SFHs, stellar masses, line EWs, and UV luminosities) of the $z\sim6$ LBGs, we fit their photometry with the BayEsian Analysis of GaLaxy sEds (\textsc{beagle}) code \citep{Chevallard2016}.
\textsc{beagle} utilizes the \citet{Gutkin2016} stellar photoionization models which were constructed by processing the updated \citet{Bruzual2003} stellar population synthesis models with isochrones computed by the PAdova and TRieste Stellar Evolution Code (\textsc{parsec}; \citealt{Bressan2012_parsec,Chen2015_parsec}) through \textsc{cloudy} \citep{Ferland2013}.

In our \textsc{beagle} fits, we employ the two-component SFH (TcSFH) parametrization described in \citetalias{Endsley2024} which allows for a very wide range of recent variations in SFR.
To summarize, the adopted TcSFH consists of a delayed-$\tau$ component (SFR $\propto t\,e^{-t/\tau_{_\mathrm{SF}}}$) and a constant SFR component.
The delayed $\tau$ component is fit to begin at least 10$^{1.35}$ ($\approx$22) Myr ago with a maximum onset time equal the age of the Universe at the fitted redshift, while the constant SFH component starts between 1--20 Myr ago (log-uniform prior for the onset times of each component) and extends to the epoch of observation.
The delayed $\tau$ component is turned off over the time interval spanned by the constant SFR component, thereby allowing for high flexibility in the recent SFH shapes in the TcSFH model fits.
The models can have strong recent downturns in SFR, strong recent upturns in SFR, or even relatively constant recent SFHs.

In \textsc{beagle}, the SFR during the constant component is parameterized by the specific SFR (sSFR) over this time period (relative to the final formed stellar mass).
This sSFR allowed to vary between 0.00001 Gyr$^{-1}$ and 1000 Gyr$^{-1}$ with a log-uniform prior.
We also allow for a wide range of $\tau_{_\mathrm{SF}}$ (1 Myr to 30 Gyr), metallicities (0.0063 to 0.5 $Z_\odot$), stellar masses (10$^{5}$ to $10^{12}$ \Msol{}), and ionization parameters (0.0001 to 0.1) adopting log-uniform priors for each.
The stellar and interstellar medium (ISM) metallicities are fixed to the same value within the \citet{Gutkin2016} models, though the effective gas-phase metallicity is altered by dust-grain depletion.
This depletion is parameterized by the dust-to-metal mass ratio, $\xi_d$, which we allow to vary between 0.1 and 0.5 (uniform prior).
Dust attention is implemented as a single screen using the SMC law \citep{Pei1992} with a log-uniform prior on the rest-frame V-band optical depth ($0.001 \leq \tau_\mathrm{V} \leq 5$).
Finally, we adopt a uniform prior on the fitted redshift in the range $z=4-8$ given our focus on $z\sim6$ LBGs.

The posterior probability distributions for each free parameter in the \textsc{beagle} fits are derived using the Bayesian \textsc{multinest} algorithm \citep{Feroz2008,Feroz2009}.
When quoting the inferred values of individual objects, we take the median of the posterior probability distribution as the fiducial value, with the 16th percentile and 84th percentile values reflecting the $\pm$1$\sigma$ uncertainty bounds.
Because of our focus on recent SFHs, it is crucial to have constraints on both H$\alpha$ and \OIIIHb{} EWs from the photometry (see \citetalias{Endsley2024}).
We therefore exclude from our sample all galaxies with $>$16\% probability of having a redshift $z>6.6$ since H$\alpha$ moves out of the NIRCam wavelength coverage at this redshift.
This removes only 12 galaxies (all in the GOODS subset), leaving a total sample of 368 $z\sim6$ LBGs that we consider in our analysis.

For a subset of our LBGs showing firm evidence of recent strong SFR downturns, we aim to quantify their recent SFHs with high precision (in context of the \citealt{Gutkin2016} models).
We therefore also consider binned (i.e., `non-parametric') SFH models with \textsc{beagle}, where it is assumed that the SFH can be described as a piecewise series of tophat functions \citep[e.g.,][]{Leja2019}.
We adopt ten time bins, the first seven of which are fixed to lookback time intervals that well sample the evolution of EWs and Balmer break amplitude following a strong burst of star formation (see \S\ref{sec:strongDownturnLBGs}): 0--3 Myr, 3--5 Myr, 5--10 Myr, 10--20 Myr, 20--30 Myr, 30--50 Myr, and 50--100 Myr.
The final three time bins implemented in our fits are evenly spaced in log lookback time between 100 Myr and $t_\mathrm{max}$, where $t_\mathrm{max}$ is set to allow (not require) star formation to begin at $z=20$.
Therefore, for spectroscopically-confirmed objects (\S\ref{sec:spectra}), $t_\mathrm{max} = t_\mathrm{Universe}(z=z_\mathrm{spec}) - t_\mathrm{Universe}(z=20)$.
Otherwise, we fit allowing for redshifts in the 1--99th percentile range from the TcSFH fit posteriors (i.e., $z_\mathrm{phot,1} - z_\mathrm{phot,99}$, and so $t_\mathrm{max} = t_\mathrm{Universe}(z=z_\mathrm{phot,99}) - t_\mathrm{Universe}(z=20)$ in the binned SFH fits.
In each of the ten time bins, the SFR is fit to a constant value with a log-uniform prior between 0.001 $M_\odot$/yr and 100 $M_\odot$/yr. 

\subsection{Inferred SED Properties of LBG Sample} \label{sec:LBGproperties}

\begin{figure}
\includegraphics[width=\columnwidth]{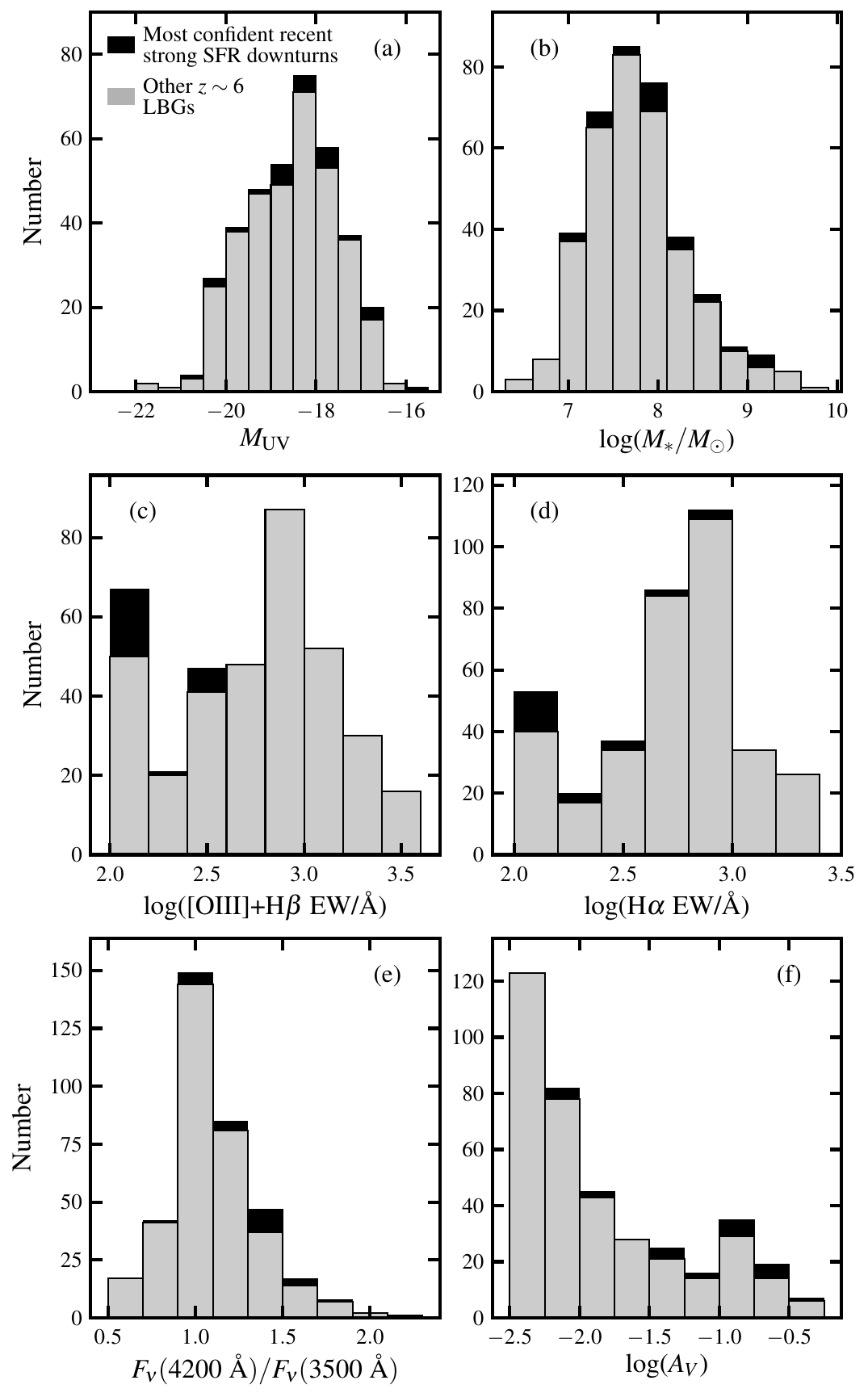}
\caption{Distribution of photometrically-inferred properties among our sample of 368 $z\sim6$ LBGs using the \textsc{beagle} TcSFH fits (\S\ref{sec:photoionization}). The subset of 24 LBGs that have most confidently experienced a recent strong downturn in SFR (\S\ref{sec:strongDownturnSelection}) are shown with a darker shading. Galaxies with inferred values outside the range of the plotted space (e.g., those with inferred EWs$<$100 \AA{}) are put in the nearest shown bin for clarity.}
\label{fig:propertyDistributions}
\end{figure}

To place our $z\sim6$ LBG sample into context, we here provide summary statistics of basic SED properties (\Muv{}, \Mstar{}, \OIIIHb{} EW, H$\alpha$ EW, \BBshort{}, and $A_V$) inferred from the \textsc{beagle} TcSFH fits to the photometry.
For galaxies in the A2744 field, we correct the absolute UV magnitudes and stellar masses for the estimated lensing magnification.
The distribution of inferred properties are illustrated in Fig. \ref{fig:propertyDistributions}.

The absolute far-UV magnitudes (rest-frame 1500 \AA{}) among the 368 $z\sim6$ LBGs span the range $-22.0 \leq \Muv{} \leq -15.8$ (thus covering a factor of $\approx$300 in UV luminosity), with a median value of $\Muv{} = -18.4$ (Fig. \ref{fig:propertyDistributions}a).
Using a characteristic UV luminosity ($L^\ast_\mathrm{UV}$) equivalent to $\Muv{} = -20.5$ \citep[e.g.,][]{Harikane2022}, our sample spans 0.013--4.0 $L^\ast_\mathrm{UV}$ and a median of 0.14 $L^\ast_\mathrm{UV}$.
The 16--84th percentile \Muv{} range spanned by our sample is $-19.6 \leq \Muv{} \leq -17.5$, corresponding to 0.06--0.4 $L^\ast_\mathrm{UV}$.

The inferred stellar masses span the range $\approx$3$\times$10$^6$ \Msol{} to $\approx$6$\times$10$^9$ \Msol{}, with a median value of $\approx$5$\times$10$^7$ \Msol{} (Fig. \ref{fig:propertyDistributions}b).
The inner 68\% interval is 2$\times$10$^7$ \Msol{} to 2$\times$10$^8$ \Msol{}.
It has been shown that the inferred stellar masses of $z\sim6$ galaxies can depend significantly on the adopted SFH prior and parametrization, particularly for objects where the emitted light is heavily dominated by recently-formed O stars (e.g., \citealt{Topping2022_REBELS,Whitler2023_z7,Narayanan2024}).
\citetalias{Endsley2024} demonstrated that, for $z\sim6$ LBGs, there is typically only an 0.1--0.3 dex offset between the stellar masses inferred from \textsc{beagle} TcSFH fits and those inferred from SED fits adopting different SFH assumptions with \textsc{beagle} and \textsc{prospector} \citep{Johnson2021}.

The inferred \OIIIHb{} and H$\alpha$ EWs of our $z\sim6$ LBG sample also span a very wide range.
We infer \OIIIHb{} EWs between 0 to 4100 \AA{}, with a median and inner 68\% range of 630 \AA{} and 110--1390 \AA{}, respectively (Fig. \ref{fig:propertyDistributions}c).
The inferred H$\alpha$ EWs span 0--2700 \AA{}, with a median and inner 68\% range of 610 \AA{} and 190--1010 \AA{} (Fig. \ref{fig:propertyDistributions}d).
Our sample thus includes a substantial number of extreme line emitters implying recent strong upturns in SFR, as well as a considerable subset of systems with very low \OIIIHb{} and H$\alpha$ EWs perhaps due to recent strong downturns in SFR (see \S\ref{sec:strongDownturnLBGs}).

Following \citet{Binggeli2019}, we quantify the Balmer break amplitude as the ratio of the flux density at rest-frame 4200 \AA{} to that at 3500 \AA{}, i.e. \BBrat{} which we shorthand as \BBshort{}.
We infer a large dynamic range of Balmer break amplitudes among our sample, spanning $0.58 \leq \BBshort{} \leq 2.16$ with a median value of \BBshort{} = 1.06 (Fig. \ref{fig:propertyDistributions}e). 
The inner 68\% range of $0.90 \leq \BBshort{} \leq 1.35$ indicates that the bulk of our $z\sim6$ LBGs show weak-to-moderate Balmer breaks.

Finally, we consider the range of dust attenuation strengths within our sample.
The median V-band optical depth of all 368 $z\sim6$ LBGs is $A_V = 0.01$ mag, suggesting typically very weak attenuation.
The full range of optical depths spans $A_V = 0.00-0.49$ mag (Fig. \ref{fig:propertyDistributions}f), indicating that our LBG selection is capable of picking up sources with strong dust attenuation.
The inner 68\% range spanned by our sample is $A_V = 0.01-0.10$ mag.

\begin{figure}
\includegraphics[width=\columnwidth]{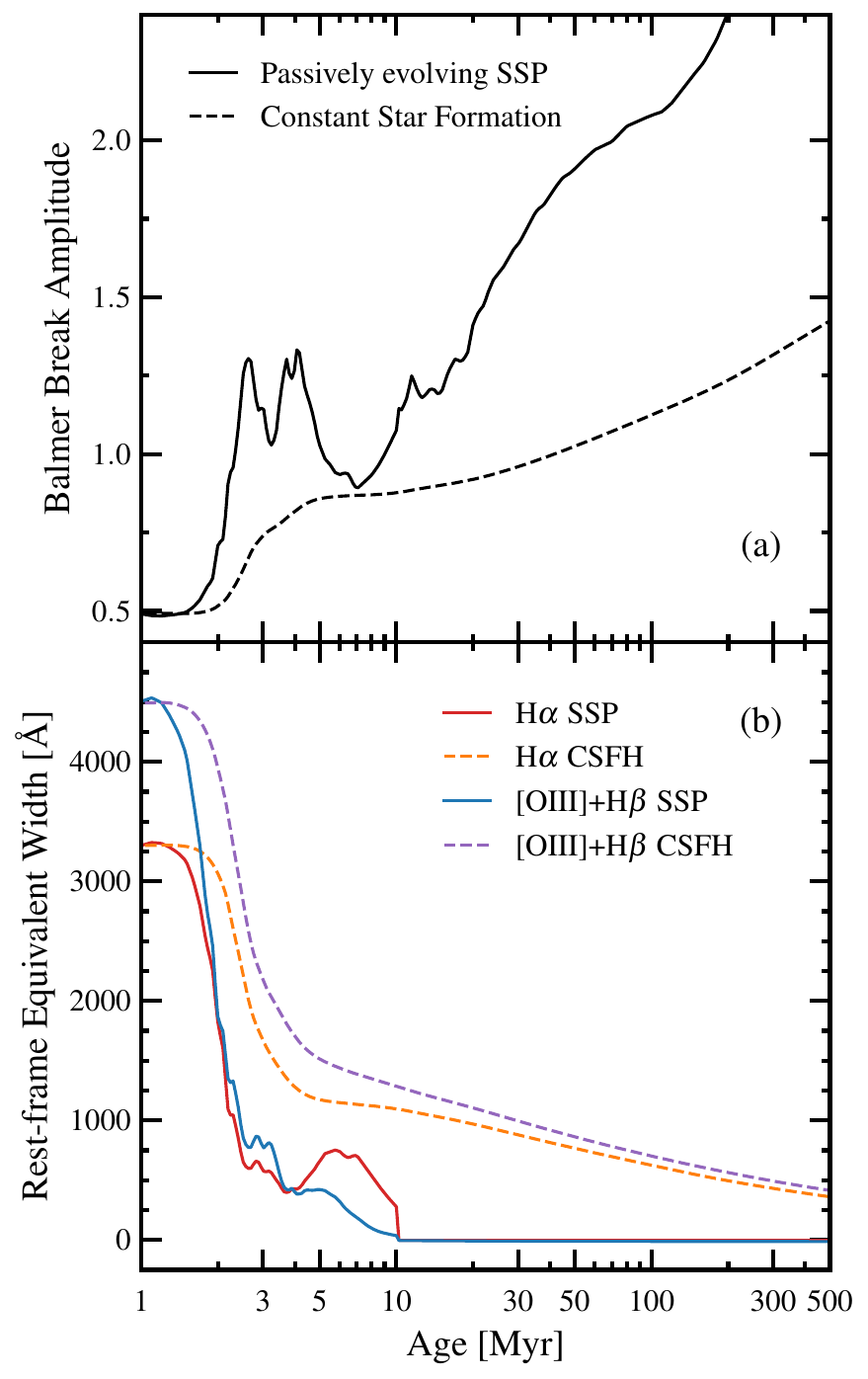}
\caption{An illustration of how the Balmer break amplitude (\BBrat{}; panel a) and rest-frame nebular line EWs (panel b) evolve with time using the single-star \citet{Gutkin2016} photoionization models. Here, we adopt moderate gas conditions ($Z = 0.1\,Z_\odot$, log $U = -2.5$) and no dust attenuation. Fig. \ref{fig:BPASS} (Appendix \ref{app:binaries}) shows how these tracks change when using binary stellar evolution models.}
\label{fig:BB_EW_evolution}
\end{figure}

\subsection{LBGs with the Most Confident Recent Strong Downturns in SFR} \label{sec:strongDownturnLBGs}

We now identify and characterize objects from our parent LBG sample that have most confidently experienced a recent strong downturn in SFR, thereby building our understanding of this recently-discovered early galaxy population.
Our procedure for selecting these galaxies from the SED fit posteriors is described in \S\ref{sec:strongDownturnSelection}.
We then discuss the inferred SED properties of these galaxies, comparing to previously-reported samples of similar objects in the literature (\S\ref{sec:strongDownturnProperties}).

\subsubsection{The Selection Criteria} \label{sec:strongDownturnSelection}

To identify galaxies with recent strong downturns in SFR, we select those where the \textsc{beagle} TcSFH posteriors imply that the average SFR over the past 3 Myr (SFR$_\mathrm{3\,Myr}$) is much lower than average SFR over the past 3--50 Myr (SFR$_\mathrm{3-50\,Myr}$).
These inferred \SFRrat{} ratios are primarily constrained by three SED properties that can be inferred from NIRCam photometry at $z\sim6$: the amplitude of the Balmer break, the \OIIIHb{} EW, and the H$\alpha$ EW.
In context of the stellar photoionization models adopted in \textsc{beagle} \citep{Gutkin2016}, the Balmer break amplitude increases rapidly after a strong burst of star formation as the light from O stars diminishes relative to that from B and A stars.
Strong Balmer breaks ($\BBshort{} \sim 1.5$) can emerge just $\approx$30 Myr after a strong burst, compared to the $\gtrsim$500 Myr timescales required under approximately constant SFR conditions (see Fig. \ref{fig:BB_EW_evolution}a).

Nebular lines also rapidly weaken following a strong burst, reaching relatively low EWs in just $\approx$3 Myr and disappearing entirely after 10 Myr (in the \citealt{Gutkin2016} models; Fig. \ref{fig:BB_EW_evolution}b).
Of course, the nebular line EWs will depend not only on the recent SFH, but also the gas conditions (e.g., metallicity and ionization parameter).
By using the \SFRrat{} posteriors from \textsc{beagle} to select candidates, we are implicitly accounting for the range of model parameters (i.e., SFH, gas conditions, dust optical depth) consistent with all available rest-UV and optical photometry.
Variations to these parameters will not only change the line EWs, but also the shape of the UV and optical continua as well as the \OIII{}/H$\alpha$ flux ratio, all of which are constrained by the NIRCam data at $z\sim6$.
We acknowledge that there are other possible interpretations for the weak \OIII{} and Balmer lines aside from strong SFR downturns.
In some cases, extremely efficient ionizing photon escape ($\fesc{} > 0.5$; see \citetalias{Endsley2024}) or strong differential dust extinction may contribute to low EWs.
In what follows, we assume that bursty SFHs are the most natural explanation \citep[e.g.,][]{Kimm2015,Ceverino2018,Ma2018_burstySFH,Barrow2020,Katz2023,Narayanan2024}.
We later demonstrate (\S\ref{sec:specDownturns}) that the NIRSpec data show no clear signatures of very high \fesc{} or strong differential extinction among our candidates with strong SFR downturns.

\begin{table*}
\centering
\caption{Properties of the 23 $z\sim6$ LBGs that have most confidently (Equations \ref{eq:SFRratCriterion} and \ref{eq:chiSqCut}) experienced a recent strong downturn in SFR. All physical properties listed here are derived from \textsc{beagle} TcSFH fits to the photometry. The three galaxies with deep NIRSpec/prism spectra (see \S\ref{sec:specDownturns}) are marked with daggers at the end of their IDs.}
\begin{tabular}{P{2.0cm}P{1.6cm}P{1.6cm}P{1.1cm}P{1.4cm}P{1.5cm}P{1.8cm}P{2.0cm}P{2.0cm}} 
\hline
\\[-20pt]
ID & RA & Dec & $\mu$ & $z$ & \Muv{} & log($M_\ast$/$M_\odot$) & log($\frac{\mathrm{SFR_{3\,Myr}}}{\mathrm{SFR_{3-50\,Myr}}}$) & $\frac{F_\nu(4200 \mathrm{\mathring{A})}}{F_\nu(3500 \mathrm{\mathring{A})}}$ \Tstrut{}\Bstrut{} \\[4pt]
\hline 
\\[-20pt]
GN-4527 & 12:36:56.927 & +62:09:07.42 & - & $5.77^{+0.12}_{-0.19}$ & $-18.9^{+0.1}_{-0.1}$ & $8.2^{+0.2}_{-0.2}$ & $-2.1^{+1.1}_{-2.5}$ & $1.40^{+0.12}_{-0.11}$\\[4pt]
GN-28474 & 12:36:52.372 & +62:11:36.96 & - & $5.51^{+0.27}_{-0.27}$ & $-18.5^{+0.2}_{-0.3}$ & $8.7^{+0.2}_{-0.2}$ & $-4.1^{+1.5}_{-1.3}$ & $1.42^{+0.08}_{-0.07}$\\[4pt]
GN-46522 & 12:36:54.192 & +62:13:19.60 & - & $5.42^{+0.06}_{-0.08}$ & $-20.2^{+0.1}_{-0.1}$ & $9.1^{+0.2}_{-0.3}$ & $-3.7^{+1.8}_{-1.6}$ & $1.33^{+0.08}_{-0.07}$\\[4pt]
GN-50244 & 12:36:47.417 & +62:13:37.18 & - & $6.07^{+0.11}_{-0.08}$ & $-20.2^{+0.1}_{-0.1}$ & $9.2^{+0.3}_{-0.3}$ & $-3.2^{+1.8}_{-1.8}$ & $1.39^{+0.12}_{-0.09}$\\[4pt]
GN-54448 & 12:37:02.494 & +62:13:59.84 & - & $5.75^{+0.09}_{-0.09}$ & $-19.2^{+0.1}_{-0.1}$ & $8.9^{+0.2}_{-0.2}$ & $-3.4^{+1.6}_{-1.4}$ & $1.78^{+0.07}_{-0.08}$\\[4pt]
GN-64741 & 12:36:38.181 & +62:17:19.80 & - & $5.81^{+0.14}_{-0.12}$ & $-18.7^{+0.1}_{-0.1}$ & $8.0^{+0.2}_{-0.2}$ & $-3.5^{+1.9}_{-1.7}$ & $1.18^{+0.12}_{-0.15}$\\[4pt]
GN-75700 & 12:36:27.257 & +62:16:24.42 & - & $5.93^{+0.18}_{-0.17}$ & $-18.2^{+0.2}_{-0.2}$ & $8.0^{+0.3}_{-0.2}$ & $-3.5^{+1.7}_{-1.6}$ & $1.41^{+0.15}_{-0.11}$\\[4pt]
GN-99551 & 12:36:46.600 & +62:14:30.87 & - & $5.90^{+0.22}_{-0.19}$ & $-17.9^{+0.2}_{-0.2}$ & $7.8^{+0.2}_{-0.2}$ & $-3.5^{+1.7}_{-1.6}$ & $1.07^{+0.15}_{-0.14}$\\[4pt]
GS-80873\textsuperscript{\textdagger} & 03:32:36.400 & $-$27:48:33.30 & - & $5.68^{+0.06}_{-0.05}$ & $-17.9^{+0.1}_{-0.1}$ & $7.5^{+0.2}_{-0.2}$ & $-2.9^{+1.9}_{-2.3}$ & $1.03^{+0.13}_{-0.10}$\\[4pt]
GS-91043 & 03:32:43.908 & $-$27:47:45.05 & - & $6.26^{+0.07}_{-0.11}$ & $-17.7^{+0.1}_{-0.1}$ & $7.3^{+0.2}_{-0.1}$ & $-3.5^{+2.2}_{-1.9}$ & $0.92^{+0.13}_{-0.07}$\\[4pt]
GS-94758 & 03:32:28.761 & $-$27:47:27.62 & - & $5.90^{+0.13}_{-0.13}$ & $-17.5^{+0.2}_{-0.2}$ & $7.9^{+0.2}_{-0.3}$ & $-3.3^{+1.7}_{-1.8}$ & $1.26^{+0.10}_{-0.13}$\\[4pt]
GS-98136 & 03:32:40.562 & $-$27:47:11.94 & - & $5.82^{+0.11}_{-0.11}$ & $-18.2^{+0.1}_{-0.1}$ & $8.1^{+0.3}_{-0.2}$ & $-3.1^{+1.8}_{-1.9}$ & $1.36^{+0.13}_{-0.11}$\\[4pt]
GS-120585 & 03:32:21.822 & $-$27:44:39.01 & - & $6.27^{+0.06}_{-0.07}$ & $-19.6^{+0.1}_{-0.1}$ & $8.4^{+0.1}_{-0.1}$ & $-3.6^{+1.6}_{-1.7}$ & $1.34^{+0.05}_{-0.04}$\\[4pt]
GS-128337 & 03:32:36.326 & $-$27:45:18.00 & - & $5.97^{+0.06}_{-0.08}$ & $-18.8^{+0.1}_{-0.1}$ & $7.9^{+0.2}_{-0.1}$ & $-3.3^{+2.3}_{-2.0}$ & $1.11^{+0.09}_{-0.08}$\\[4pt]
GS-129132 & 03:32:35.564 & $-$27:45:21.71 & - & $5.32^{+0.18}_{-0.17}$ & $-18.0^{+0.1}_{-0.1}$ & $8.0^{+0.3}_{-0.1}$ & $-3.4^{+1.7}_{-1.8}$ & $1.50^{+0.09}_{-0.09}$\\[4pt]
GS-141885 & 03:32:38.287 & $-$27:46:17.49 & - & $6.09^{+0.06}_{-0.06}$ & $-20.5^{+0.1}_{-0.1}$ & $9.2^{+0.2}_{-0.2}$ & $-3.3^{+1.7}_{-1.8}$ & $1.30^{+0.09}_{-0.07}$\\[4pt]
A2744-20941\textsuperscript{\textdagger} & 00:14:28.830 & $-$30:22:43.42 & 1.5 & $5.59^{+0.07}_{-0.23}$ & $-18.5^{+0.1}_{-0.1}$ & $8.3^{+0.3}_{-0.1}$ & $-3.4^{+1.7}_{-1.8}$ & $1.50^{+0.08}_{-0.10}$\\[4pt]
A2744-21195 & 00:14:16.487 & $-$30:22:49.02 & 3.7 & $5.64^{+0.04}_{-0.04}$ & $-17.8^{+0.1}_{-0.0}$ & $7.5^{+0.1}_{-0.1}$ & $-3.6^{+1.8}_{-1.8}$ & $0.98^{+0.06}_{-0.04}$\\[4pt]
A2744-22062 & 00:14:17.888 & $-$30:23:06.83 & 4.5 & $5.77^{+0.05}_{-0.05}$ & $-16.9^{+0.1}_{-0.1}$ & $6.9^{+0.1}_{-0.1}$ & $-3.6^{+2.1}_{-1.8}$ & $0.91^{+0.08}_{-0.06}$\\[4pt]
A2744-23813 & 00:14:21.947 & $-$30:23:48.00 & 13.2 & $5.67^{+0.07}_{-0.06}$ & $-16.5^{+0.1}_{-0.1}$ & $7.7^{+0.2}_{-0.2}$ & $-4.1^{+1.4}_{-1.3}$ & $1.54^{+0.04}_{-0.04}$\\[4pt]
A2744-23814 & 00:14:21.925 & $-$30:23:48.01 & 13.9 & $5.63^{+0.11}_{-0.10}$ & $-15.8^{+0.1}_{-0.1}$ & $7.3^{+0.1}_{-0.3}$ & $-4.0^{+1.5}_{-1.4}$ & $1.44^{+0.05}_{-0.06}$\\[4pt]
A2744-26071\textsuperscript{\textdagger} & 00:14:16.814 & $-$30:24:13.28 & 2.3 & $5.65^{+0.04}_{-0.06}$ & $-18.8^{+0.1}_{-0.1}$ & $8.7^{+0.1}_{-0.1}$ & $-2.8^{+1.6}_{-2.2}$ & $1.52^{+0.04}_{-0.04}$\\[4pt]
A2744-44335 & 00:14:18.201 & $-$30:23:21.36 & 5.7 & $5.62^{+0.03}_{-0.03}$ & $-16.8^{+0.0}_{-0.0}$ & $6.9^{+0.1}_{-0.1}$ & $-3.6^{+2.1}_{-1.8}$ & $0.87^{+0.08}_{-0.05}$\\[4pt]
\hline
\end{tabular}
\label{tab:properties}
\end{table*}

When selecting $z\sim6$ LBGs with strong recent downturns in SFR, we seek to identify those with \SFRrat{} ratios that are approximately the reciprocal of extreme emission line galaxies (EELGs).
It has been shown that EELGs have very recently ($\lesssim$3 Myr ago) experienced a strong upturn in SFR, leading to an SED dominated by early-type O stars \citep[e.g.,][]{Whitler2023_z7,Boyett2024_jades}.
EELGs at $z\sim6$ are typically defined as those with \OIIIHb{} EW$\gtrsim$1200 \AA{} or H$\alpha$ EW$\gtrsim$800 \AA{} \citep[e.g.,][]{Smit2015,RobertsBorsani2016,Endsley2021_OIII,Boyett2024_jades}.
Upon generating a large number of TcSFH models with moderate ionized gas conditions (log $U = -2.5$ and $Z = 0.1\ Z_\odot$), we find that such extremely high \OIIIHb{} and H$\alpha$ EWs nearly always require $\SFRrat{} > 5$.
Therefore, we identify the most confident candidates with recent strong downturns in SFR as those with $>$90\% probability of having $\SFRrat{} < 0.2$, i.e.:
\begin{equation} \label{eq:SFRratCriterion}
    P\left(\mathrm{SFR_{3\,Myr}}/\mathrm{SFR_{3-50\,Myr}} < 0.2\right) > 0.9.
\end{equation}

In our analysis of recent SFHs, a significant source of systematic uncertainty is the potential impact of binary stellar evolution.
We qualitatively consider the effect binary evolution may have on our SFH inferences using \textsc{bpass}+\textsc{cloudy} photoionization models (see Appendix \ref{app:binaries} and Fig. \ref{fig:BPASS}).
These \textsc{bpass} models predict a similar evolution of the Balmer break amplitude in the $\gtrsim$10 Myr following a strong burst.
However, as expected, the predicted evolution in EWs is very different.
The \textsc{bpass} models predict that an SSP with moderate gas conditions ($Z \approx 0.1\ Z_\odot$ and log $U \approx -2.5$) will continue to show moderate-to-high EW \OIII{} emission ($\gtrsim$500 \AA{}) even after $\approx$10 Myr, while the \citet{Gutkin2016} models predict that such EWs persist for only $\approx$3 Myr.
This means that it is even harder for the \textsc{bpass} models to explain galaxies with surprisingly low EWs at fixed Balmer break amplitude without resorting to strong recent downturns in SFR.
On this basis alone, we generally expect that if we were to instead use the \textsc{bpass} photoionization models, the posteriors would reflect even stronger confidence of low \SFRrat{} among our candidates.
However, \textsc{bpass} does generally predict higher \OIII{}/H$\alpha$ ratios compared to the \citet{Gutkin2016} models (due to the harder radiation fields), which would also change the inferred metallicities and ionization parameters.
This may have a non-trivial impact on the inferred recent SFH shapes.
We leave future work to directly quantify how implementing binary stellar evolution models would alter our conclusions on the recent SFHs of $z\sim6$ LBGs.

We now test how well we can identify $z\sim6$ galaxies with strong recent declines in SFR when employing the criterion in Eq. \ref{eq:SFRratCriterion}.
To do so, we run mock $z=6$ SSPs through \textsc{beagle} TcSFH fits using the photometric filter sets of the GOODS and A2744 fields.
For each field we fit two mock SEDs, one an 8 Myr-old SSP and the other a 30 Myr-old SSP, both generated with the \citet{Gutkin2016} photoionization models using log $U = -2.5$, $Z = 0.1\ Z_\odot$, and no dust attenuation.
During the fits, the photometric noise is set to the same value in every filter for simplicity, and normalized such that S/N=10 in F115W.
In every case, the \textsc{beagle} TcSFH fit yields $>$97.5\% probability that the mock SED satisfies $\SFRrat{} < 0.2$.
This demonstrates that our SED-fitting approach will indeed confidently identify $z\sim6$ galaxies with very recent ($\sim$5--10 Myr ago) and somewhat recent ($\sim$30 Myr ago) strong downturns in SFR, assuming moderate S/N photometry ($\gtrsim$10$\sigma$) and that the downturns immediately follow a strong burst.

\begin{figure*}
\includegraphics[width=\textwidth]{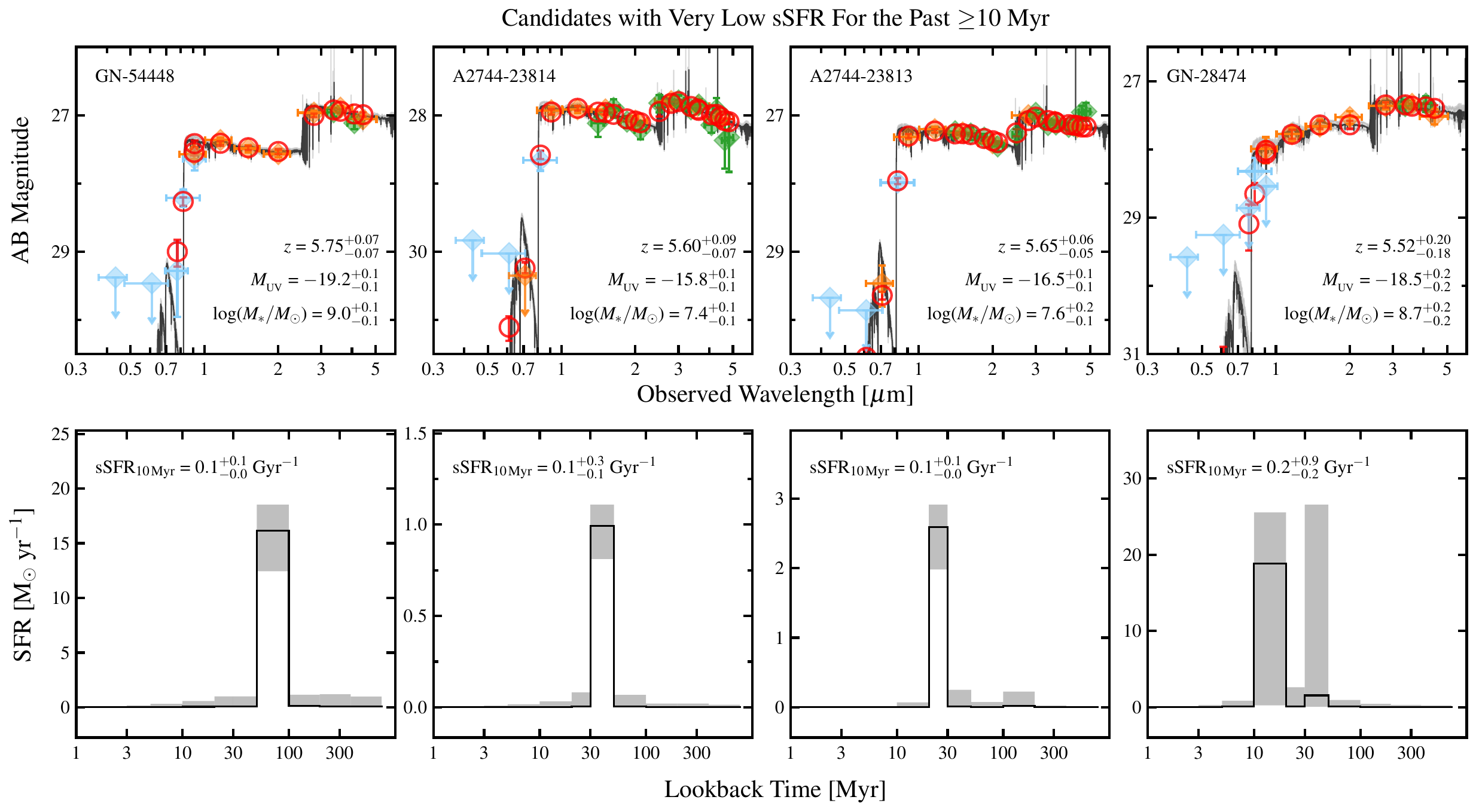}
\caption{The four $z\sim6$ LBGs that have most confidently experienced very low sSFRs for the past $\geq$10 Myr, similar to JADES-GS-z7-01-QU \citep{Looser2024}. Top panels show the \textsc{beagle} binned (i.e., `non-parametric') SFH SED fits to the ACS (blue), NIRCam broad-band (orange), and NIRCam medium-band (green) photometry. 2$\sigma$ upper limits are shown, with the model synthetic photometry as red circles. The bottom panels show the inferred SFHs, where the posteriors imply very low sSFR for the past $\approx$30--50 Myr for some. Here, properties are inferred from the binned SFH fit posteriors and corrected for lensing where appropriate (the apparent magnitudes are not corrected for lensing).}
\label{fig:veryLowsSFR10}
\end{figure*}

We also verify that the TcSFH models yield a much better fit to the photometric data relative to constant SFH (CSFH) models.
These CSFH fits are also run with \textsc{beagle} adopting the same setup as the TcSFH fits, except using a single CSFH component with allowed age between 1 Myr and the age of the Universe at the sampled redshift.
When adopting the A2744 filter set, we find that the TcSFH fits to the 8 Myr-old and 30 Myr-old mock SSPs described above yield best-fitting $\chi^2$ values at least 14 below that of the best-fitting CSFH $\chi^2$ values (i.e., $\chi^2_\mathrm{CSFH} - \chi^2_\mathrm{TcSFH} \geq 14$).
When adopting the GOODS filter set, we find $\chi^2_\mathrm{CSFH} - \chi^2_\mathrm{TcSFH} \geq 5.7$ for the two mock SSPs; the lower $\chi^2$ difference is due to the fewer number of filters in the GOODS field relative to A2744.
Because we aim to develop a rather strict selection, we require (in addition to Eq. \ref{eq:SFRratCriterion}) that our most confident candidates with strong recent downturns in SFR have $\chi^2$ differences at least $\approx$1.5$\times$ these values from the mock tests:
\begin{equation} \label{eq:chiSqCut}
    \begin{split}
        & \bullet \ \chi^2_\mathrm{CSFH} - \chi^2_\mathrm{TcSFH} > 20 \ \parbox{7em}{for A2744} \\
        & \bullet \ \chi^2_\mathrm{CSFH} - \chi^2_\mathrm{TcSFH} > 9 \ \parbox{7em}{for GOODS.}
    \end{split}
\end{equation}

The combined selection criteria (Equations \ref{eq:SFRratCriterion} and \ref{eq:chiSqCut}) yield 23 galaxies that have most confidently experienced a recent strong downturn in SFR from our parent sample of 368 $z\sim6$ LBGs.
Sixteen of these candidates come from the GOODS subset and seven come from the A2744 subset.
The coordinates and inferred properties (redshift, \Muv{}, \Mstar{}, \SFRrat{}, and \BBshort{}) of all 23 candidates are reported in Table \ref{tab:properties}.

We also checked the $\SFRrat{}$ posteriors from the binned SFH \textsc{beagle} fits for all 23 candidates.
Our intent is to ensure that the TcSFH models have sufficient flexibility in recent SFH shapes to robustly identify cases of confident strong recent SFR downturns.
For nearly all galaxies (20/23), the binned SFH fits also return $>$90\% probability of $\SFRrat{} < 0.2$.
For the remaining three galaxies, the binned SFH posteriors clearly favor a strong SFR downturn 5--10 Myr ago, but with a slight increase in SFR over the past 3 Myr to power moderate EW line emission (one such case is discussed in detail in \S\ref{sec:GS-80873}).
Overall, we conclude that we can use the (simpler and faster) TcSFH fits to identify confident cases of strong recent SFR downturns, bearing in mind that the binned SFH fits yield more precise constraints on the timing of the downturn (in context of the adopted stellar population synthesis models) as intended.

\subsubsection{SED Properties of Candidates} \label{sec:strongDownturnProperties}

Among the 23 $z\sim6$ candidates with strong recent SFR downturns, eight have strong inferred Balmer breaks ($1.4 < \BBshort{} < 1.8$) from the TcSFH fits.
All eight galaxies show NIRCam photometry implying extremely weak nebular line emission, with maximum inferred EWs of 90 \AA{} for \OIIIHb{} and 150 \AA{} for H$\alpha$.
In half of these systems (GN-28474, GN-54448, A2744-23813, A2744-23814), there is simply no convincing evidence of any rest-optical nebular line emission from the long-wavelength NIRCam colors (see Fig. \ref{fig:veryLowsSFR10}).
This results in acceptable \textsc{beagle} TcSFH and binned SFH model solutions with Balmer line absorption and no \OIII{} emission, similar to the $z=7.3$ `mini-quenched' galaxy JADES-GS-z7-01-QU \citep[][see also \citealt{Weibel2024}]{Looser2024}.
From the photometry alone, we typically infer that their sSFRs over the past 10 Myr were $\lesssim\,0.2$ Gyr$^{-1}$ (Fig. \ref{fig:veryLowsSFR10}), very similar to the spectroscopic sSFR$_\mathrm{10\,Myr} \lesssim\,0.1$ Gyr$^{-1}$ limit for JADES-GS-z7-01-QU \citep{Looser2024}.

Among these four candidates with very low recent sSFRs, GN-54448 stands out as having a very prominent inferred Balmer break with $\BBshort{} \approx 1.8$, much stronger than JADES-GS-z7-01-QU ($\BBshort{} \approx 1.4$; \citealt{Looser2024}).
Accordingly, the binned SFH fit implies that GN-54448 has likely experienced very little star formation for the past $\approx$50 Myr, considerably longer than the $\approx$20 Myr timescale inferred for JADES-GS-z7-01-QU \citep{Looser2024}.
Both GN-54448 and JADES-GS-z7-01-QU have fairly large inferred stellar masses of $\Mstar{} \approx (0.5-1)\times10^9\,\Msol{}$.
But we also identify two systems at much lower stellar mass ($\Mstar{} \approx 3\times10^7\,\Msol{}$) that appear to have experienced negligible star formation for the past $\approx$20--30 Myr, more similar to the $z=5.2$ galaxy identified in \citet{Strait2023}.
These two clumps in our sample (A2744-23813 and A2744-23814) are very nearby one another ($\approx$0.3\arcsec{} separation) and are both highly magnified ($\mu \approx 13.5$), so they may occupy the same dark matter halo.
Given existing lensing constraints for the foreground clusters, A2744-23813 and A2744-23814 are very unlikely to be the same multiply-imaged clump (L. Furtak, private communication).
The inferred SFHs of these two clumps suggest they formed in rather quick succession ($\Delta t \sim 10$ Myr).
Deep NIRSpec follow-up is needed to more accurately and precisely infer the timescale over which all four LBGs shown in Fig. \ref{fig:veryLowsSFR10} have experienced very low sSFRs.

Our LBG sample also includes seven galaxies with confident strong recent SFR downturns and weak Balmer breaks ($0.85 < \BBshort{} < 1.15$; see Table \ref{tab:properties}).
These systems simply have much lower \OIIIHb{} and H$\alpha$ EWs than expected for the young light-weighted ages implied by their continua (see \citealt{Endsley2023_CEERS,Endsley2024}), implying a relative deficit of early-type O stars compared to slightly less massive stars dominating the continuum. 
Two of these galaxies (GS-80873 and GS-128337) were also identified as systems with weak Balmer breaks and surprisingly weak nebular lines via a different selection technique in \citetalias{Endsley2024}.
From the binned SFH fits, all seven galaxies are inferred to have far lower sSFRs over the past 3 Myr ($\approx$0.1--1 Gyr$^{-1}$) compared to that 5-20 Myr ago ($\approx$30--50 Gyr$^{-1}$).
In \S\ref{sec:GS-80873}, we provide a detailed analysis of one such galaxy with deep NIRSpec spectra.

We now consider how the UV luminosities and stellar masses of the 23 candidates with strong recent SFR downturns compare with that of the full parent $z\sim6$ LBG sample (\S\ref{sec:LBGproperties}).
The 23 candidates have UV luminosities spanning approximately two orders of magnitude ($-20.5 \leq \Muv{} \leq -15.8$) with a median $\Muv{} = -18.2$, nearly equivalent to that of the parent sample ($\Muv{} = -18.4$).
However, because our selection criteria (Equations \ref{eq:SFRratCriterion} and \ref{eq:chiSqCut}) incorporate a confidence interval cut from the SED fit posteriors, it is more challenging for fainter LBGs (with generally lower S/N photometry) to satisfy this selection.
Therefore, the median UV luminosity of \textit{all} galaxies with strong recent declines in SFR across our parent sample (including those not satisfying the selection criteria) is likely lower than the value quoted above.
Indeed, we demonstrate that the combined SED constraints of all galaxies in our sample support this conclusion in \S\ref{sec:SFRratDistnsMuv}.

Galaxies with recent strong declines in SFR will be relatively faint in their UV continuum at fixed stellar mass (see Fig. \ref{fig:SSPevolution}).
Indeed, the median stellar mass of the 23 candidates is $\Mstar{} = 1.0\times10^8\ \Msol{}$, a factor of two higher than the typical stellar mass of the full LBG parent sample despite the nearly equal median \Luv{}.
Nonetheless, we again emphasize that the identification bias towards brighter systems will also impact the stellar mass distribution of the most confident subset with $\SFRrat{} < 0.2$.
The typical stellar mass of all galaxies with strong recent declines in SFR across our parent LBG sample (including those not satisfying the confidence cuts of Equations \ref{eq:SFRratCriterion} and \ref{eq:chiSqCut}) is thus likely lower than the value quoted above.
Future work will be required to estimate the selection completeness of galaxies with strong recent downturns in SFR as a function of, e.g. UV luminosity, stellar mass, and the time since the most recent `burst' (see \S\ref{sec:discussion}).

\section{Spectroscopic Validation of Photometric SED Inferences} \label{sec:spectra}

A primary goal of this paper is to utilize the photometrically-inferred properties of our full $z\sim6$ LBG sample to statistically constrain the distribution of recent SFHs among early galaxies.
In this section, we utilize public NIRSpec/prism spectroscopy to test and ultimately validate the photometric SED inferences before conducting statistical analyses on the full sample in \S\ref{sec:analysis}.
We first describe the spectroscopic sample and how we measure galaxy properties (redshifts, line fluxes, line EWs, and Balmer breaks) from the spectra in \S\ref{sec:specMethods}.
Then we compare these spectroscopic measurements with the photometrically-inferred SED properties (EWs, Balmer break amplitudes, recent SFH shapes) in \S\ref{sec:specComparison}.
Finally, in \S\ref{sec:specDownturns}, we analyze spectroscopic data on three candidates with strong recent downturns in SFR (\S\ref{sec:strongDownturnLBGs}), demonstrating that the spectra are fully consistent with that physical interpretation.

\subsection{The Spectroscopic Sample} \label{sec:specMethods}

To determine which of our $z\sim6$ LBGs were observed with NIRSpec/prism, we begin by cross-matching the LBG coordinates with the central MSA shutter positions of each target.
Twenty-nine NIRSpec/prism targets had a central MSA position within 0.7 arcsec of one of our LBGs.
For each of those 29 targets, we visually check the on-sky placement of each MSA shutter configuration to ensure that 1) the shutters actually overlap significantly with the light profile of the nearby LBG, and 2) that the shutters are not strongly contaminated by light from another object on the sky.
A total of four objects are removed by these visual cuts, resulting in a final sample of 25 $z\sim6$ LBGs with usable NIRSpec/prism observations.
Twelve lie in A2744 and thirteen lie in GOODS.

The prism spectra reveal detections of the \OIII{} and H$\alpha$ emission lines for all 25 galaxies.
In the majority of targets (19/25), the spectra are also sufficiently deep to show a Ly$\alpha$ break (or Ly$\alpha$ emission in a few cases).
To measure spectroscopic redshifts, we perform Gaussian fits to the \OIII{} and H$\alpha$ lines (a double Gaussian in the case of \OIII{}, fixing the wavelength separation to that expected of the unresolved doublet) and compute the redshift from each line using vacuum wavelengths.
For each galaxy, we then compute the inverse variance-weighted mean from the two redshift measurements.

We measure the Balmer break amplitude as well as fluxes and EWs of various lines (\OII{}$\lambda\lambda$3727,3729, \NeIII{}$\lambda$3869, \OIII{}, H$\beta$, and H$\alpha$) directly from the NIRSpec spectra.
Line fluxes are computed by directly integrating the continuum-subtracted spectra, and we assume negligible contribution from \NII{} emission when calculating the H$\alpha$ flux \citep{Cameron2023_jades,Sanders2024}.
To measure the continuum flux densities, we perform separate power law fits (using \textsc{emcee}; \citealt{ForemanMackey2013_emcee}) to the rest-UV data (using the range 1300--3500 \AA{}) and rest-optical data (using the range 4000--8000 \AA{}) after masking regions impacted by emission lines.

These 25 spectroscopic targets cover a large dynamic range in SED properties.
The UV luminosities and stellar masses span $\approx$2 dex, the \OIIIHb{} and H$\alpha$ EWs include very small ($\lesssim$100 \AA{}) and extreme ($\sim$2500 \AA{}) values, and the Balmer break amplitudes span $\BBshort{} \approx 0.7-1.8$.
The median UV luminosity and stellar mass of these 25 galaxies are slightly (0.2 dex) higher than that of the parent LBG sample.

\begin{figure}
\includegraphics[width=\columnwidth]{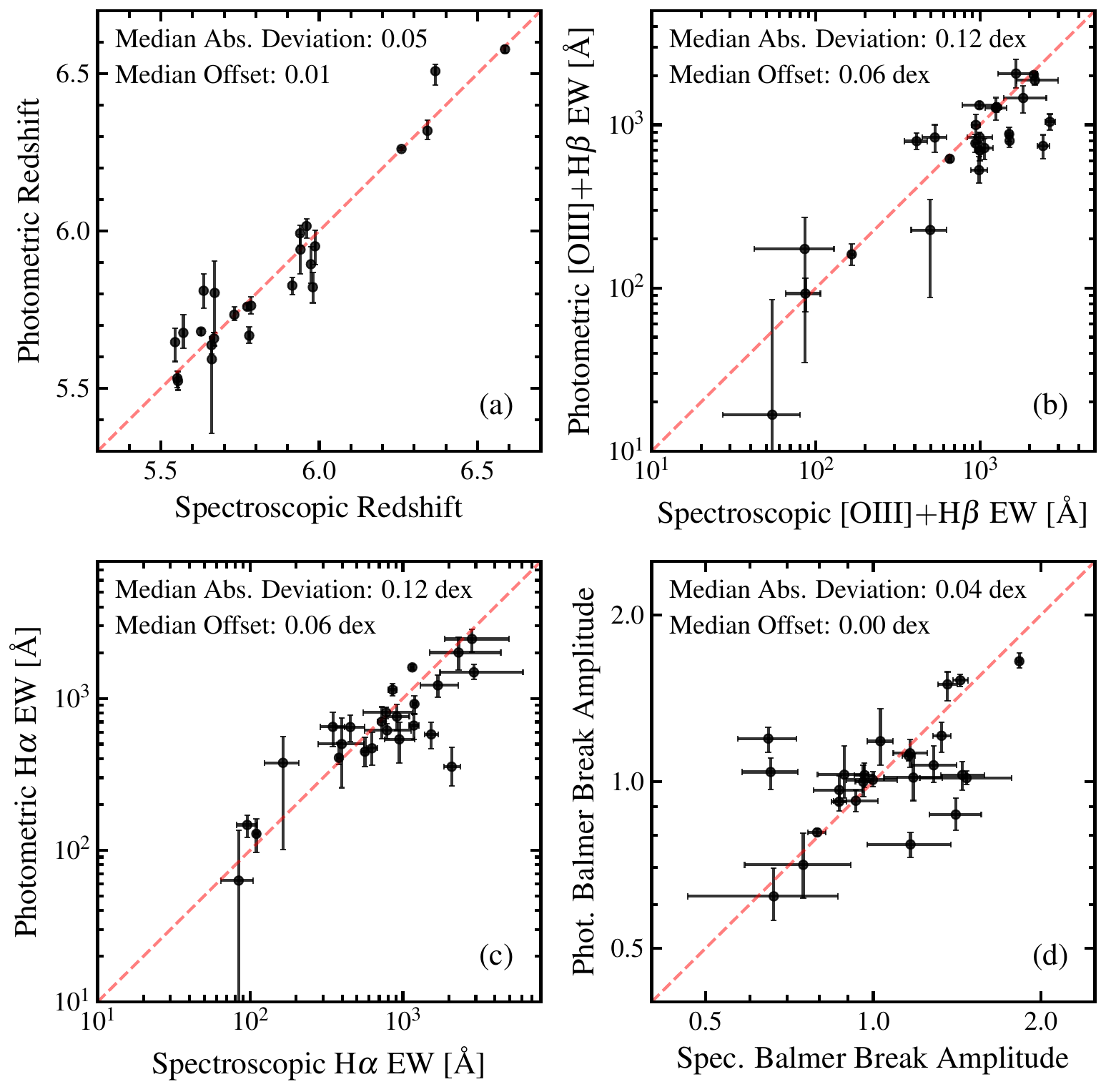}
\caption{A comparison of the redshifts, \OIIIHb{} EWs, H$\alpha$ EWs, and Balmer break amplitudes measured from NIRSpec prism spectra (x-axis) and that inferred from \textsc{beagle} TcSFH fits to only the photometry (y-axis). The red dashed lines show the one-to-one relation, and the errorbars show the 16--84th percentile uncertainty range. There is no significant systematic offset in any of these four properties.}
\label{fig:specComparison}
\end{figure}

\subsection{Comparison with Photometrically-Inferred Properties} \label{sec:specComparison}

All 25 spectroscopically-targeted galaxies have measured redshifts in the range $\zspec{} = 5.545-6.587$ (see Fig. \ref{fig:specComparison}a), fully consistent with our intended selection of $z\sim6$ LBGs at $z<6.6$ (where H$\alpha$ remains covered by NIRCam).
The median offset between the spectroscopic and photometric redshifts is 0.01 and the median absolute deviation is 0.05.
This implies that there is no significant systematic offset in the photometric redshifts for our $z\sim6$ LBG sample.

We also find that there are no strong systematic offsets in the photometrically-inferred \OIIIHb{} EWs, H$\alpha$ EWs, nor Balmer break amplitudes (Fig. \ref{fig:specComparison}b--d).
The median offset of these three properties are, respectively, 0.06, 0.06, and 0.00 dex.
Moreover, their median absolute deviations are all modest at 0.12 dex, 0.12 dex, and 0.04 dex, respectively.
Accordingly, we conclude that the SED properties inferred from the \textsc{beagle} TcSFH fits are generally robust among our LBG sample.

\begin{figure}
\includegraphics[width=\columnwidth]{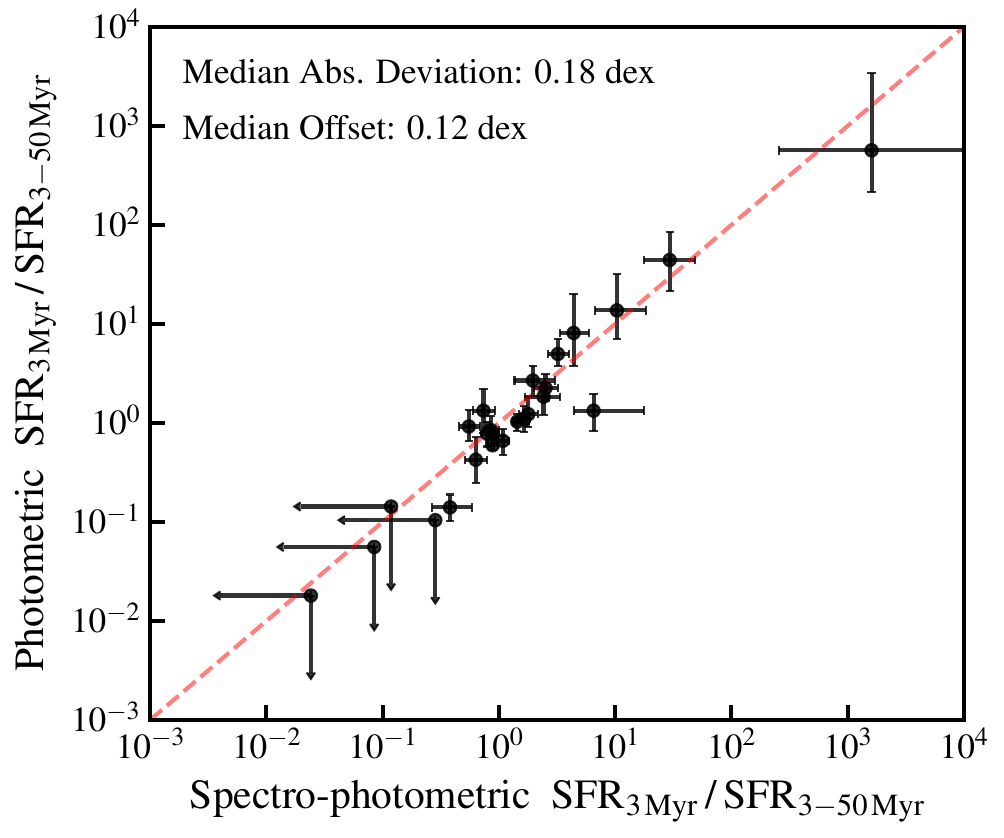}
\caption{Similar to Fig. \ref{fig:specComparison}, except comparing the inferred recent SFH shape between the photometry-only and spectro-photometric TcSFH fits. We show 84th percentile upper limits where appropriate. We conclude that our photometry-only inferences on the recent SFHs of $z\sim6$ LBGs are reasonably robust.}
\label{fig:specComparison_SFRrat}
\end{figure}

The inferred recent SFHs of our LBGs are largely constrained by the \OIIIHb{} EWs, the H$\alpha$ EWs, and Balmer break amplitudes inferred from the photometric colors.
Because our photometric inferences on all three of these SED properties appear robust against spectroscopic measurements, we thus expect our photometric inferences on the recent SFHs to also be generally robust.
We explicitly test this by re-running the \textsc{beagle} TcSFH fits for the 25 NIRSpec/prism targeted galaxies, folding in spectroscopic measurements along with the photometry.
In these spectro-photometric \textsc{beagle} fits, we input the spectroscopic redshifts along with the \OII{}, \NeIII{}, \OIII{}, H$\beta$, and H$\alpha$ integrated fluxes. 
We then compare the \SFRrat{} ratios inferred from the photometry-only fits to those inferred from these spectro-photometric fits.

There is a strong correlation between the recent SFHs inferred from the photometry alone and those inferred when folding in the spectroscopic measurements (Fig. \ref{fig:specComparison_SFRrat}).
For all 12 galaxies where the photometry-only fits infer a recently-declining SFH ($\SFRrat{} < 1$), the spectro-photometric fits also infer recently-declining SFHs within the inner 68\% posterior uncertainties.
Similar consistency is found for those where the photometry-only fits infer a recently-rising SFH ($\SFRrat{} > 1$).
Considering all twenty-five $z\sim6$ LBGs with NIRSpec/prism spectra, we find a small (0.12 dex) median offset between the spectro-photometric and photometry-only inferences on \SFRrat{}, with modest median absolute deviation (0.18 dex) as well.
Ultimately, we conclude that our photometric inferences on the recent SFHs of $z\sim6$ galaxies are reasonably robust.

\subsection{Confirmation of Recent Strong SFR Downturns} \label{sec:specDownturns}

To date, there have been three $z>5$ galaxies with spectroscopic signatures indicating a strong recent downturn in SFR \citep{Strait2023,Looser2024,Weibel2024}.
Here, we double that sample.
There are three new spectroscopically-confirmed galaxies in our $z\sim6$ LBG sample where we inferred a strong recent SFR downturn from the photometry (GS-80873, A2744-20941, and A2744-26071; see \S\ref{sec:strongDownturnLBGs}).
The following sub-sub-sections discuss each of the three galaxies in turn, first describing why the photometry alone confidently implied a strong recent SFR downturn, then demonstrating why the spectra fully support this conclusion.

\begin{figure*}
\includegraphics[width=\textwidth]{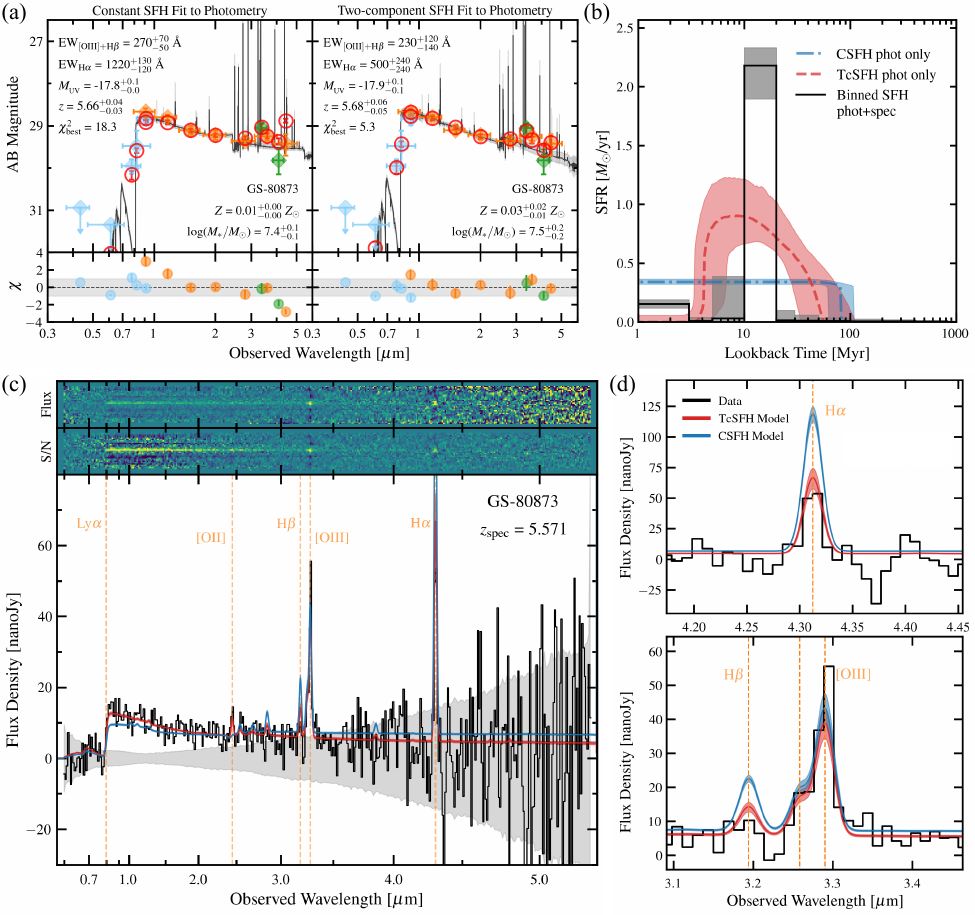}
\caption{Empirical evidence for a recent strong decline in SFR in a very UV-faint $\zspec{} = 5.571$ galaxy, GS-80873. \textbf{(a)} \textsc{beagle} SED fits to the ACS and NIRCam photometry similar to Fig. \ref{fig:veryLowsSFR10}. The left panel shows a fit with a pure constant SFH (CSFH) model, while the right panel shows the fit using a two-component SFH (TcSFH) model. The lower panels show the significance with which the model synthetic photometry differs from the measurements in terms of $\chi \equiv \left(f_\mathrm{meas} - f_\mathrm{model}\right)/\sigma_\mathrm{meas}$. \textbf{(b)} The inferred SFHs from the two photometry-only model fits (dashed lines), as well as a binned SFH fit including line flux measurements from the spectra (black). \textbf{(c)} The NIRSpec prism spectrum. Top panels show the 2D flux and S/N images, while the optimally-extracted 1D spectrum is shown in the bottom panel. We mark the expected wavelength positions of a few emission features at the spectroscopic redshift. \textbf{(d)} Zoomed in panels on the H$\alpha$ line (top), as well as the H$\beta$ line and \OIII{} doublet (bottom). In panels (c) and (d), we also overlay the \textsc{beagle} TcSFH and CSFH SED models which have been fit to the rest-optical line flux measurements from the spectra in addition to the photometry.}
\label{fig:GS80873}
\end{figure*}

\begin{figure*}
\includegraphics[width=\textwidth]{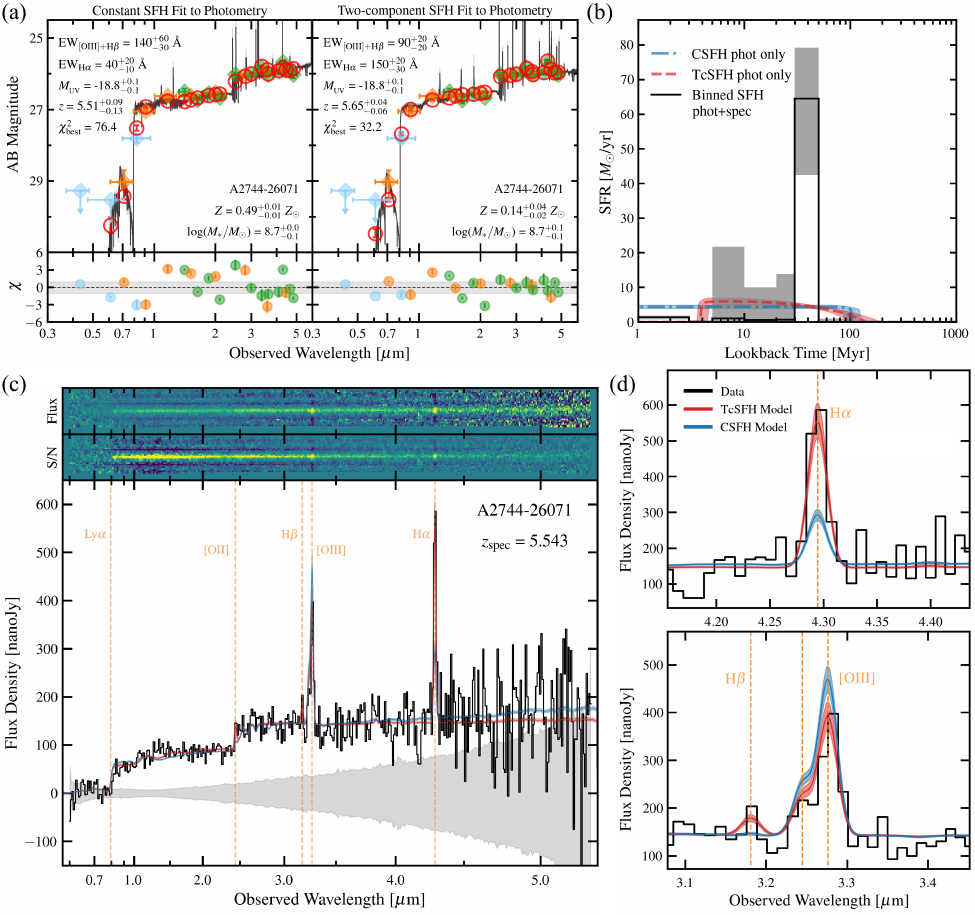}
\caption{Same format as Fig. \ref{fig:GS80873} except for the galaxy A2744-26071 at $\zspec{} = 5.543$. The inferred properties are (where appropriate) corrected for gravitational lensing, but the apparent magnitudes and flux densities are given in observed units.}
\label{fig:A2744_26071}
\end{figure*}

We have considered whether the NIRSpec data on these three galaxies imply that their relatively weak nebular emission lines are instead due to extremely efficient Lyman-continuum (LyC) photon escape ($\fesc{} > 0.5$) or strong differential extinction between the stellar continuum and nebular emission.
None show evidence that either of these alternative explanations are more likely than recent strong declines in SFR.
We test the feasibility of extremely-high \fesc{} solutions using the strong anti-correlation between \fesc{} and the observed far-UV continuum slope reported at $z\sim0.3$ \citep{Chisholm2022}.
From the NIRSpec spectra, we measure far-UV (1300--1800 \AA{} rest-frame) continuum slopes in the range $-0.2 \leq \beta_\mathrm{FUV} \leq -1.9$ for all three galaxies.
\citet{Chisholm2022} found that all $\approx$50 $z\sim0.3$ galaxies in the Low-$z$ Lyman Continuum Survey with similarly red far-UV slopes ($-2 \leq \beta_{FUV} \leq 0$) have $\fesc{} < 0.1$, far less than required to significantly decrease line EWs.
Moreover, as shown in Figures \ref{fig:GS80873}--\ref{fig:A2744_20941}d, the single-screen SMC dust attenuation law that we adopt in our fits is sufficient to explain the measured H$\alpha$/H$\beta$ flux ratios within uncertainties for all three galaxies. 

\subsubsection{GS-80873} \label{sec:GS-80873}

GS-80873 is a very UV-faint ($\Muv{} = -17.9\pm0.1$) galaxy in the GOODS-S field with an estimated stellar mass of $\Mstar{} \approx 3\times10^7$ \Msol{}.
NIRSpec/prism data from the JOF survey (MSA ID 30080873) has now confirmed that this galaxy lies at $\zspec{} = 5.571 \pm 0.001$, in good agreement with its estimated photometric redshift ($\zphot{} \approx 5.7$).
We first describe why the photometry alone so confidently implied a strong recent SFR downturn before demonstrating the additional constraining power of the NIRSpec data.

GS-80873 shows a weak photometric excess in F335M implying a small \OIIIHb{} EW of $\sim$250 \AA{}, yet no sign of a Balmer break (F200W$-$F277W = $-$0.14$\pm$0.14 mags) implying a relatively young light-weighted age.
There are two ways that \textsc{beagle} can try to reproduce weak \OIII{} at young light-weighted ages using standard CSFH models.
The first is by pushing the models to extremely high ionization parameter, where nebular continuum emission is strengthened while nebular metal line emission is weakened.
Such a scenario is ruled out by the very blue far-to-near UV slope ($\beta = -2.7\pm0.1$) measured from the NIRCam photometry, since strong nebular continuum emission substantially reddens the UV slope (see, e.g., \citealt{Byler2017,Topping2022_blueSlopes}).

The \textsc{beagle} CSFH models instead resort to extremely low metallicities ($\approx$0.01 $Z_\odot$) to try reproducing the low \OIIIHb{} EW and weak Balmer break seen from GS-80873.  
In such models, Balmer lines are strengthened by low stellar opacity (see \citetalias{Endsley2024} for a quantitative discussion) implying that there should be high-EW H$\alpha$ emission ($\approx$1200 \AA{}; see Fig. \ref{fig:GS80873}a).
This would result in a substantial photometric excess in F444W at $z\sim6$.
However, this prediction is inconsistent with the photometry at the $\approx$3$\sigma$ level (Fig. \ref{fig:GS80873}a).
Moreover, extremely low metallicities significantly redden the UV continuum via stronger nebular emission, compounding the tension between the data and CSFH model.

The photometry of GS-80873 is instead well explained by a TcSFH model wherein the SFR rapidly declined in the last $\approx$5--10 Myr (Fig. \ref{fig:GS80873}a,b).
In these models, nebular emission is greatly weakened by the relative dearth of early-type O stars.
The TcSFH model nonetheless maintains a young continuum due to the surviving late-type O and B stars that formed $\approx$10--50 Myr ago.
One of the most striking differences between the two \textsc{beagle} fits is that the TcSFH models predict an H$\alpha$ EW of $\approx$500 \AA{}, much weaker than implied by the CSFH fits ($\approx$1200 \AA{}).
The TcSFH model also successfully reproduces the very blue shape of the UV continuum.

We now use the NIRSpec data to verify that the TcSFH model truly yields a better match to the UV through optical SED.
The \OIIIHb{} and H$\alpha$ EWs measured from the prism spectrum are $480^{+110}_{-100}$ \AA{} and $400^{+160}_{-130}$ \AA{}, respectively.
Only the TcSFH model fit correctly predicts the relatively low H$\alpha$ EW (Fig. \ref{fig:GS80873}a),
Crucially, the young light-weighted age we infer from the photometry is supported by the prism spectrum which shows no sign of a Balmer break (\BBshort{} = $0.89^{+0.09}_{-0.10}$). 
Upon fitting the combined spectro-photometric measurements with \textsc{beagle}, we find that CSFH models simply fail to reproduce the \JWST{} data while TcSFH models yield acceptable fits (Fig. \ref{fig:GS80873}c,d).
Specifically, the best-fitting CSFH model ($\chi^2 = 157$) yields a 6$\sigma$ offset in H$\alpha$ flux, a $\approx$2.5$\sigma$ offset in H$\beta$ flux, and a systematic $\approx$1.5$\sigma$ offset in the far-UV continuum (rest-frame 1250--2000 \AA{}) across $\approx$35 spectral pixels.
On the other hand, the best-fitting TcSFH model ($\chi^2 = 26$) reproduces the measured H$\alpha$ and H$\beta$ fluxes within $\approx$1$\sigma$ and yields a much better match to the continuum as well.
It is therefore clear that the data support a strong recent decline in SFR over standard CSFH models.

The spectro-photometric binned SFH fit offers more precise constraints on the recent SFH of GS-80873 (see Fig. \ref{fig:GS80873}b).
In this highly-flexible \textsc{beagle} SED fit, GS-80873 underwent a very strong burst of star formation 10--20 Myr ago (average SFR = 2.2$^{+0.3}_{-0.1}$ \Msol{}/yr over this time bin) and has remained far less active since (SFR$\sim$0.1 \Msol{}/yr).
The average SFR over the past 3 Myr is precisely constrained to 0.15$^{+0.04}_{-0.03}$ \Msol{}/yr by the nebular line flux measurements.
Moreover, the lack of a Balmer break limits the average SFR between 20--50 Myr ago to $<$0.1 \Msol{}/yr (84th percentile limit).
Overall, this far more flexible SFH fit further supports the interpretation that GS-80873 experienced a strong recent downturn in SFR.

\subsubsection{A2744-26071} \label{sec:A2744-26071}

A2744-26071 is an intrinsically UV-faint ($\Muv = -18.8 \pm 0.1$) Lyman-break galaxy in the A2744 field that is moderately lensed by the foreground clusters ($\mu = 2.3$). 
The UNCOVER survey targeted this galaxy with the NIRSpec/prism (MSA ID 12065), yielding $\zspec{} = 5.543 \pm 0.001$ consistent with the photometric redshift from \textsc{beagle} ($\zphot{} \approx 5.6$).
Again, we first explain how the photometry alone confidently implied a strong recent SFR downturn before demonstrating that the NIRSpec data support this conclusion.

A2744-26071 is detected at $>$10$\sigma$ in every NIRCam broad and medium band, affording exceptionally stringent SED constraints.
The medium-band photometry reveal a prominent Balmer break, as well as weak but statistically significant excesses due to \OIIIHb{} and H$\alpha$ implying EWs of $\approx$50--150 \AA{}.
The only way the \textsc{beagle} CSFH models come even relatively close to simultaneously matching the very low EWs and moderately strong Balmer break is by pushing to the upper bound of the allowed prior space in ionization parameter (log $U \approx -1$) and metallicity ($Z \approx 0.5 Z_\odot$).
But even then, the best-fitting CSFH model yields $>$3$\sigma$ offsets in several rest-optical bands (see Fig. \ref{fig:A2744_26071}a) ultimately resulting in $\chi^2_\mathrm{best} = 76.4$ across the 23 fitted data points.
The TcSFH model provides a much better match to the data with $\chi^2_\mathrm{best} = 32.2$ and $<$1.5$\sigma$ offsets in all rest-optical bands (Fig. \ref{fig:A2744_26071}a).
Such models are able to reproduce the moderate Balmer break from the buildup of A and late-type B stars, as well as the low-EW optical line emission from the dearth of surviving early-type O stars. 

Upon folding in the line flux and redshift constraints from the NIRSpec spectra, we again find that the CSFH models greatly struggle to reproduce the data (best-fitting $\chi^2 = 103$) while the TcSFH models provide far more acceptable solutions (best-fitting $\chi^2 = 51$; Fig. \ref{fig:A2744_26071}c,d). 
While the CSFH models can successfully match the measured continuum and \OIII{} flux, they under-predict the measured H$\alpha$ flux by at least $\approx$5$\sigma$ (Fig. \ref{fig:A2744_26071}c,d). 
This is largely due to the fact that the CSFH models must push to extremely high ionization parameters (log $U \approx -1$) to come even relatively close to the very low measured \OIII{} and H$\alpha$ EWs (82$\pm$17 \AA{} and 96$\pm$15 \AA{}, respectively).
The TcSFH model, however, can reproduce all of the data with a more moderate ionization parameter (log $U = -1.4^{+0.3}_{-0.5}$) by allowing for a strong decline in SFR $\approx$5--10 Myr ago (Fig. \ref{fig:A2744_26071}b).
This recent downturn in SFR removes many of the O stars that would power strong nebular line and continuum emission in the CSFH scenario.

From the spectro-photometric binned SFH fit, we infer that A2744-26071 experienced a dramatic burst of star formation 30--50 Myr ago (average SFR = 65$^{+15}_{-23}$ \Msol{}/yr over this time bin; see Fig. \ref{fig:A2744_26071}b).
The very weak emission lines imply a far lower SFR$_\mathrm{3\,Myr} = 1.5^{+0.4}_{-0.7}$ \Msol{}/yr in this \textsc{beagle} fit.
Again, we conclude that the NIRSpec/prism data support a strong recent downturn in SFR for this galaxy.

\begin{figure*}
\includegraphics[width=\textwidth]{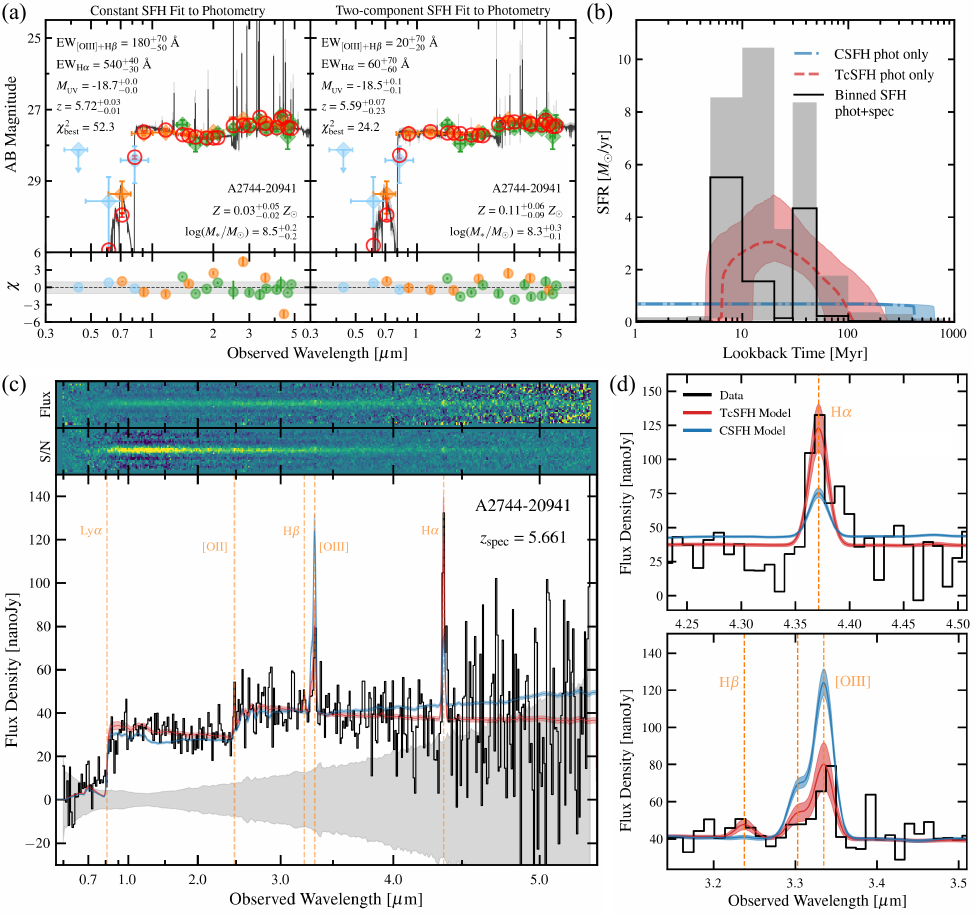}
\caption{Same format as Fig. \ref{fig:A2744_26071} except for the galaxy A2744-20941 at $\zspec{} = 5.661$.}
\label{fig:A2744_20941}
\end{figure*}

\subsubsection{A2744-20941} \label{sec:A2744-20941}

A2744-20941 is very similar to the galaxy discussed in the previous sub-sub-section (A2744-26071), but is $\approx$1--1.5 mag fainter (in apparent magnitude), in part due to its weaker lensing magnification ($\mu = 1.5$).
This galaxy is UV-faint (intrinsic $\Muv{} = -18.5\pm0.1$) and moderately massive ($\Mstar{} \approx 3\times10^8$\ \Msol{}), showing clear signatures of a strong Balmer break ($\BBshort{} \approx 1.5$) and very weak \OIIIHb{} and H$\alpha$ emission (EWs$\lesssim$100 \AA{}) from the photometry (Fig. \ref{fig:A2744_20941}a).
The NIRSpec/prism data confirm that this galaxy lies at $\zspec{} = 5.661 \pm 0.003$ via the detection of the Ly$\alpha$ break, a Balmer break, the \OIII{} line, and the H$\alpha$ line.

As was the case with A2744-26071, the CSFH model fit to the photometry yields considerable tension with the data ($\chi^2_\mathrm{best} = 52$ over 23 data points), due in part to $\approx$4--5$\sigma$ offsets in two bands (Fig. \ref{fig:A2744_20941}a). 
The photometry is far more consistent with TcSFH models ($\chi^2_\mathrm{best} = 24$) wherein the SFR fell substantially within the last $\approx$5--10 Myr (Fig. \ref{fig:A2744_20941}a,b).

The \textsc{beagle} CSFH models also fail to reproduce the combined spectro-photometric measurements of A2744-20941 ($\chi^2_\mathrm{best} = 87$), yielding an $\approx$3$\sigma$ offset in the H$\alpha$ flux, and a $\approx$3$\sigma$ offset in the \OIII{} flux, as well as noticeable offsets in the predicted UV and optical continuum (Fig. \ref{fig:A2744_20941}c,d).
The TcSFH model fits, however, successfully match all \JWST{} data on this source $\chi^2_\mathrm{best} = 31$).
The spectro-photometric binned SFH fit again supports the conclusion that A2744-20941 recently experienced a strong decline in SFR (Fig. \ref{fig:A2744_20941}b) with log($\SFRrat{}$) = $-2.1 \pm 0.7$.

\section{Analysis} \label{sec:analysis}

We have now demonstrated that deep NIRSpec/prism data validate our photometric inferences on the SEDs and recent SFHs of $z\sim6$ LBGs (\S\ref{sec:spectra}).
In this section, we utilize the full statistical constraining power of our $z\sim6$ LBG sample ($N_\mathrm{gal} = 368$) to better understand SFHs at early cosmic epochs.
We first analyze the H$\alpha$ to far-UV luminosity ratios among our sample (\S\ref{sec:Ha_UV}) and then infer the distribution of recent SFH shapes as a function of UV luminosity (\S\ref{sec:SFRratDistnsMuv}).

\begin{figure*}
\includegraphics[width=\textwidth]{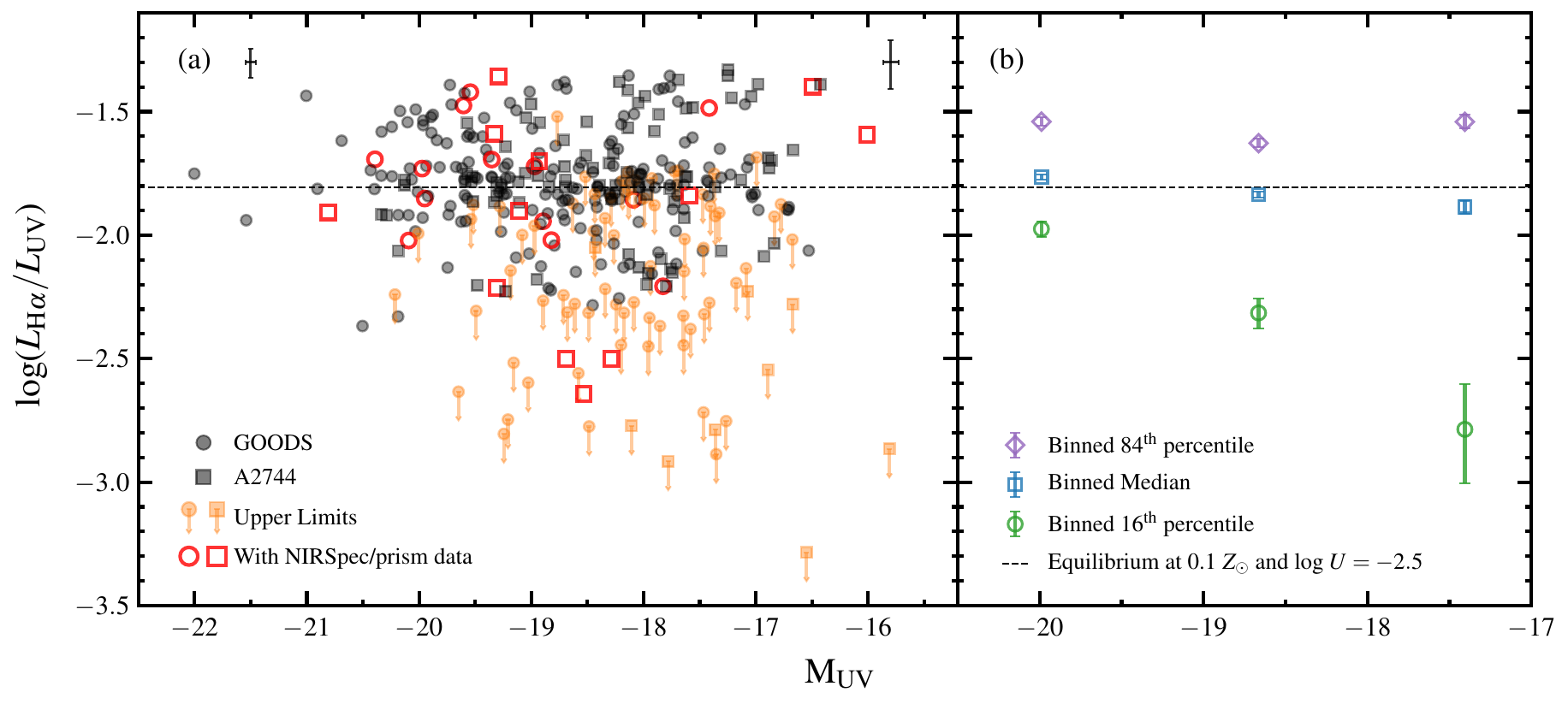}
\caption{Relation between the dust-corrected H$\alpha$ to far-UV continuum luminosity ratio (\Lrat{}) and absolute UV magnitude among our sample of 368 $z\sim6$ LBGs. Panel (a) shows the inferred values of each galaxy, while panel (b) shows the median, 16th, and 84th percentile values on \Lrat{} in three \Muv{} bins. Typical uncertainties on \Lrat{} for bright ($\Muv{} < -19.5$) and faint ($\Muv{} > -18$) objects without upper limits on $L_\mathrm{H\alpha}$ (see text) are shown as black errorbars in the upper left and right, respectively, of panel (a).}
\label{fig:Lrat_Muv}
\end{figure*}

\subsection{The \texorpdfstring{$L_\mathrm{H\alpha}$}{LHa}/\texorpdfstring{$L_\mathrm{UV}$}{Luv} Ratios of \texorpdfstring{$z\sim6$}{z ~ 6} LBGs} \label{sec:Ha_UV}

Our goals in this sub-section are twofold.
First, we check whether a subset of our $z\sim6$ LBGs exhibit extremely low H$\alpha$ to far-UV luminosity ratios ($\Lrat{} \lesssim 0.001$), as seen in local dwarf galaxies that appear to be caught between burst events \citep[e.g.,][]{Emami2019}.
Second, we test whether very low \Lrat{} values are substantially more common at the UV-faint end of our $z\sim6$ LBG sample.
Such an \Muv{}-dependent trend is expected from bursty SFHs because strong downturns in SFR rapidly decrease \Luv{} and, to a stronger degree, $L_\mathrm{H\alpha}$.

We infer the \Lrat{} ratios of all 368 $z\sim6$ LBGs using the posteriors of the \textsc{beagle} TcSFH fits. 
Both the far-UV continuum luminosities and H$\alpha$ luminosities are corrected for dust using the optical depths inferred from the \textsc{beagle} fits.
For the galaxies with NIRSpec prism data, we utilize their TcSFH fits that fold in the spectroscopic redshift and line flux measurement constraints to correct for dust attenuation in a consistent manner across the full sample.
For those without prism spectra, we adopt upper limits on the H$\alpha$ luminosity if the inferred H$\alpha$ EW is less than 100 \AA{} (84th percentile credible interval limit) or if the inner 68th percentile credible interval uncertainty on $L_\mathrm{H\alpha}$ is greater than 0.7 dex.
We have verified that our conclusions change negligibly if we instead use the photometry-only fits for all LBGs.

Within our $z\sim6$ LBG sample, we infer H$\alpha$ to far-UV luminosity ratios spanning at least 2 orders of magnitude from $\logLrat{} = -1.3$ to $\logLrat{} < -3.3$ (see Fig. \ref{fig:Lrat_Muv}a).
Nineteen LBGs show very low luminosity ratios of $\logLrat{} \leq  -2.5$, including three systems with NIRSpec/prism measurements.
Such low \Lrat{} ratios have been measured in several local dwarf ($\Mstar{} < 10^8\ \Msol{}$) galaxies, implying that these nearby low-mass objects experience strong changes in SFR on very short timescales ($\lesssim$30 Myr; e.g., \citealt{Emami2019}).
We note that \citet{Asada2024} also report very low \Lrat{} values among their sample of 57 $z\sim5-6$ LBGs from independent \JWST{} data, lending support to our findings.

We now explore how the distribution of \Lrat{} changes with UV luminosity at $z\sim6$.
It is clear from Fig. \ref{fig:Lrat_Muv}a that there are several LBGs with high luminosity ratios ($\logLrat{} \gtrsim -1.5$) at both the UV-bright ($\Muv{} < -19.5$) and very UV-faint ($\Muv{} > -18$) end of our sample.
There also appears to be a relatively high fraction of galaxies with low luminosity ratios ($\logLrat{} \lesssim -2.2$) at the faint end.
To directly quantify how the distribution of \Lrat{} changes with \Muv{}, we utilize Monte Carlo sampling to account for the uncertainties associated with each object. 
That is, for each of the 368 galaxies, we extract 1000 realizations of their inferred \Lrat{} and \Muv{} from the \textsc{beagle} TcSFH fit posterior.
We then sort the galaxies into three \Muv{} bins: UV-bright ($\Muv < -19.5$; $N_\mathrm{gal} = 76$), UV-faint ($-19.5 < \Muv{} < -18$; $N_\mathrm{gal} = 179$), and very UV-faint ($\Muv{} > -18$; $N_\mathrm{gal} = 113$).
Within each bin, we compute the median, 16th, and 84th percentile values on \Lrat{} (called $R_{50}$, $R_{16}$, and $R_{84}$ respectively for clarity below) for each of the 1000 realizations.
From those 1000 realizations, we subsequently compute the median, 16th, and 84th percentile values on $R_{50}$, $R_{16}$, and $R_{84}$. 

The median \Lrat{} ratio of our $z\sim6$ LBG sample decreases significantly with UV luminosity  (see Fig. \ref{fig:Lrat_Muv}b).
Among the UV-bright ($\langle\Muv{}\rangle = -20.0$) subset, we obtain a median $\logLrat{} = -1.76\pm0.01$ while we infer $\logLrat{} = -1.89\pm0.03$ among the very UV-faint ($\langle\Muv{}\rangle = -17.4$) subset.
This is the opposite direction of what is expected from metallicity effects alone.
Because both UV luminosity and metallicity generally increase with stellar mass (e.g., \citealt{Curti2024}; \citetalias{Endsley2024}), we expect UV-bright $z\sim6$ galaxies to overall be more metal enriched than very UV-faint galaxies.
This effect alone would have resulted in lower \Lrat{} among brighter galaxies since more metals result in higher stellar opacity to ionizing photons.
Specifically, the \citet{Gutkin2016} models predict equilibrium values (constant SFR for $>$100 Myr) of $\logLrat{} = -1.76$, $-1.81$, and $-1.82$ for $Z/Z_\odot = $ 0.01, 0.1, and 0.5, respectively, assuming a moderate ionization parameter (log $U = -2.5$).
As explained above, the decreasing \Lrat{} towards fainter UV luminosities is an expected outcome of very bursty SFHs.

We also find that there is a very strong \Muv{} dependence on the 16th percentile value of \Lrat{} (Fig. \ref{fig:Lrat_Muv}b).
While UV-bright $z\sim6$ galaxies generally only modestly scatter down to $\logLrat{} = -1.98\pm0.03$, the very UV-faint galaxies frequently exhibit ratios extending nearly 1 dex lower to $\logLrat{} = -2.79^{+0.18}_{-0.22}$.
To quantify this trend another way, we apply the same Monte Carlo sampling of the \textsc{beagle} posteriors to compute the fraction of galaxies with very low luminosity ratios ($\logLrat{} < -2.5$) in each \Muv{} bin.
We find that this fraction increases dramatically towards fainter \Muv{}, with $4\pm1$ percent of UV-bright galaxies showing $\logLrat{} < -2.5$ compared to $20\pm2$ percent of very UV-faint galaxies.
That is a $\approx$5$\times$ increase in the fraction of galaxies with very low \Lrat{} over only a $\approx$1 order of magnitude decrease in UV luminosity.
This is consistent with strong recent SFR downturns becoming much more common at the faint-end of the UV luminosity function.

Our findings here support the conclusions of \citetalias{Endsley2024} that the \Muv{}-dependent trends on the \OIIIHb{} and H$\alpha$ EW distributions at $z\sim6$ cannot be explained by metallicity effects alone.
\citetalias{Endsley2024} instead argued that the strong weakening of \OIIIHb{} towards fainter \Muv{} yet approximately fixed H$\alpha$ EW was a signature of bursty SFHs.
Our results on \Lrat{} here further demonstrate that the \Muv{}-dependence on nebular emission among $z\sim6$ LBGs cannot be explained solely by metallicity, and are instead consistent with expectations of bursty SFHs.
Of course, several other physical factors can modify H$\alpha$ and far-UV luminosities including the IMF, binary stellar evolution, stellar rotation, ionization parameter, and ionizing photon escape (see references in \citealt{Rezaee2023}). 
Our goal in this sub-section was to demonstrate that the \Lrat{} ratios among our full LBG sample are consistent with expectations of bursty SFHs, bearing in mind that other probes (like the EW distribution trends described in \citetalias{Endsley2024}) must be considered as well to make relatively firm conclusions.

\begin{figure}
\includegraphics[width=\columnwidth]{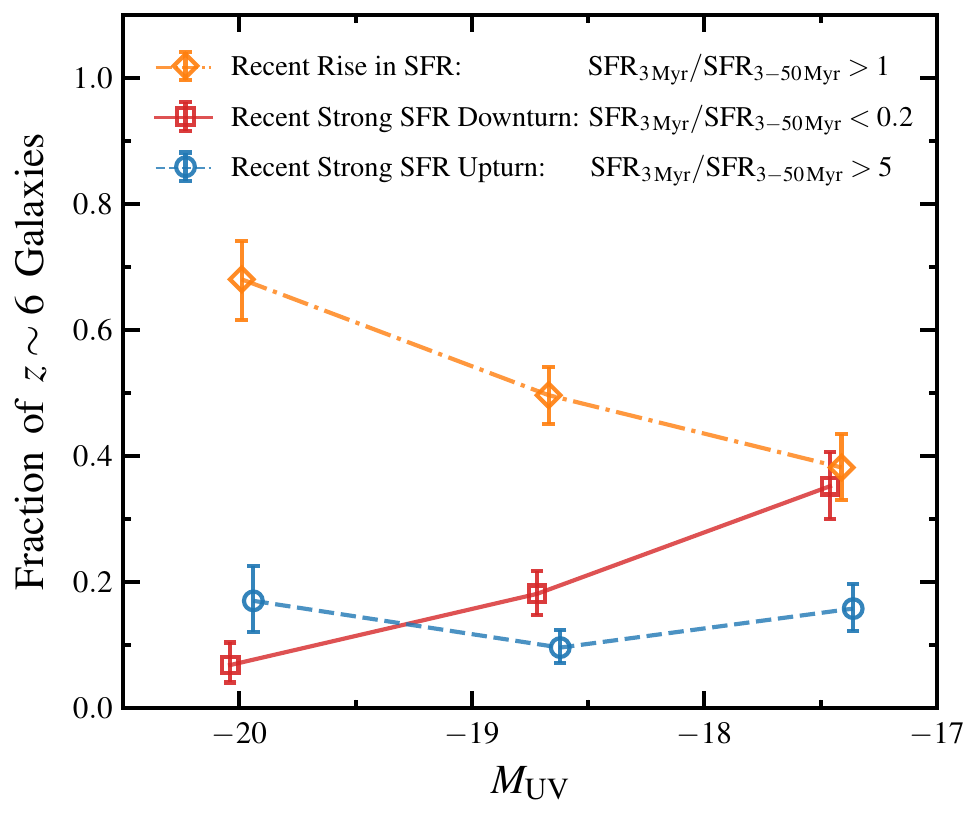}
\caption{The connection between recent SFH and far-UV luminosity at $z\sim6$ constrained by our sample of 368 LBGs. Very UV-faint ($\langle\Muv{}\rangle = -17.4$) galaxies in our sample are far more likely to have recently experienced a strong SFR downturn compared to their UV-bright counterparts ($\langle\Muv{}\rangle = -20.0$). The fraction of galaxies with strong recent upturns is fairly constant ($\approx$10--15\%) across the UV luminosity range of our sample. We note that these fractions do not sum to unity in a given \Muv{} bin because the strong upturn fraction is a subset of the rising fraction, and none of the sub-populations shown here reflect those with moderate recent downturns. See Fig. \ref{fig:SFRratPDFandCDF} for more detailed constraints on how the distribution of recent SFH shapes changes with \Muv{}.}
\label{fig:dutyCycle}
\end{figure}

\subsection{The Distribution of Recent SFH Shapes as a Function of UV Luminosity} \label{sec:SFRratDistnsMuv}

Here, we directly infer how the recent SFH shapes (quantified via \SFRrat{}) of $z\sim6$ galaxies change with UV luminosity using the posteriors from the \textsc{beagle} TcSFH fits.
As in the previous sub-section we use the spectro-photometric fits for the 25 objects with NIRSpec/prism data; we have ensured that our conclusions remain unchanged if we instead use the photometry-only fits for all.
We first infer the fraction of our LBGs undergoing strong recent SFR downturns ($\SFRrat{} < 0.2$) and strong recent SFR upturns ($\SFRrat{} > 5$) as a function of UV luminosity, using the same three \Muv{} bins as in \S\ref{sec:Ha_UV}.
These fractions are calculated by employing an established Bayesian formalism (\citealt{Schenker2014,Boyett2022_OIII}; \citetalias{Endsley2024}).
This formalism accounts for the uncertainty in \SFRrat{} for each individual object (as estimated by the TcSFH posteriors), as well as Poisson uncertainties imposed by the total number of objects in each \Muv{} bin.
The statistical constraining power afforded by our large LBG sample reveals several significant trends.

The fraction of $z\sim6$ galaxies with strong recent downturns in SFR ($\SFRrat{} < 0.2$) increases substantially towards lower UV luminosities within our sample (see Fig. \ref{fig:dutyCycle}).
In the UV-bright bin ($\langle \Muv{} \rangle = -20.0$), the fraction of galaxies with $\SFRrat{} < 0.2$ is inferred to be very small at $7\pm3$ percent.
This fraction then increases to $18^{+4}_{-3}$ percent in the UV-faint bin ($\langle \Muv{} \rangle = -18.7$), and $35^{+6}_{-5}$ percent in the very UV-faint bin ($\langle \Muv{} \rangle = -17.4$).
Therefore, we find an approximately five-fold increase in the fraction of galaxies with strong recent SFR downturns over only a factor of $\approx$10 difference in UV luminosity.
This result is consistent with the \Muv{}-dependent trend in the fraction of galaxies with very low \Lrat{} (\S\ref{sec:Ha_UV}).

The fraction of $z\sim6$ galaxies with a recent strong rise in SFR ($\SFRrat{} > 5$) is instead inferred to remain fairly constant over the \Muv{} range of our sample.
In all three UV luminosity bins, this fraction is consistent with $\approx$10--15\% within the uncertainties (Fig. \ref{fig:dutyCycle}).
This is again qualitatively consistent with our finding that the fraction of $z\sim6$ LBGs with high \Lrat{} is roughly constant with \Muv{} (\S\ref{sec:Ha_UV}).

We also constrain the fraction of galaxies undergoing a recently rising SFH to any degree (i.e., $\SFRrat{} > 1$) as a function of UV luminosity.
In the UV-bright bin ($\langle \Muv{} \rangle = -20.0$), approximately two-thirds ($68\pm6$ percent) of $z\sim6$ LBGs have experienced a recent rise in SFR.
This fraction declines significantly towards lower UV luminosities, reaching $38^{+6}_{-5}$ percent in our very UV-faint bin ($\langle \Muv{} \rangle = -17.4$; Fig. \ref{fig:dutyCycle}).
We therefore infer that less than half of very UV-faint $z\sim6$ LBGs have a higher SFR over the past 3 Myr relative to their average SFR of the past 3--50 Myr.

\begin{figure}
\includegraphics[width=\columnwidth]{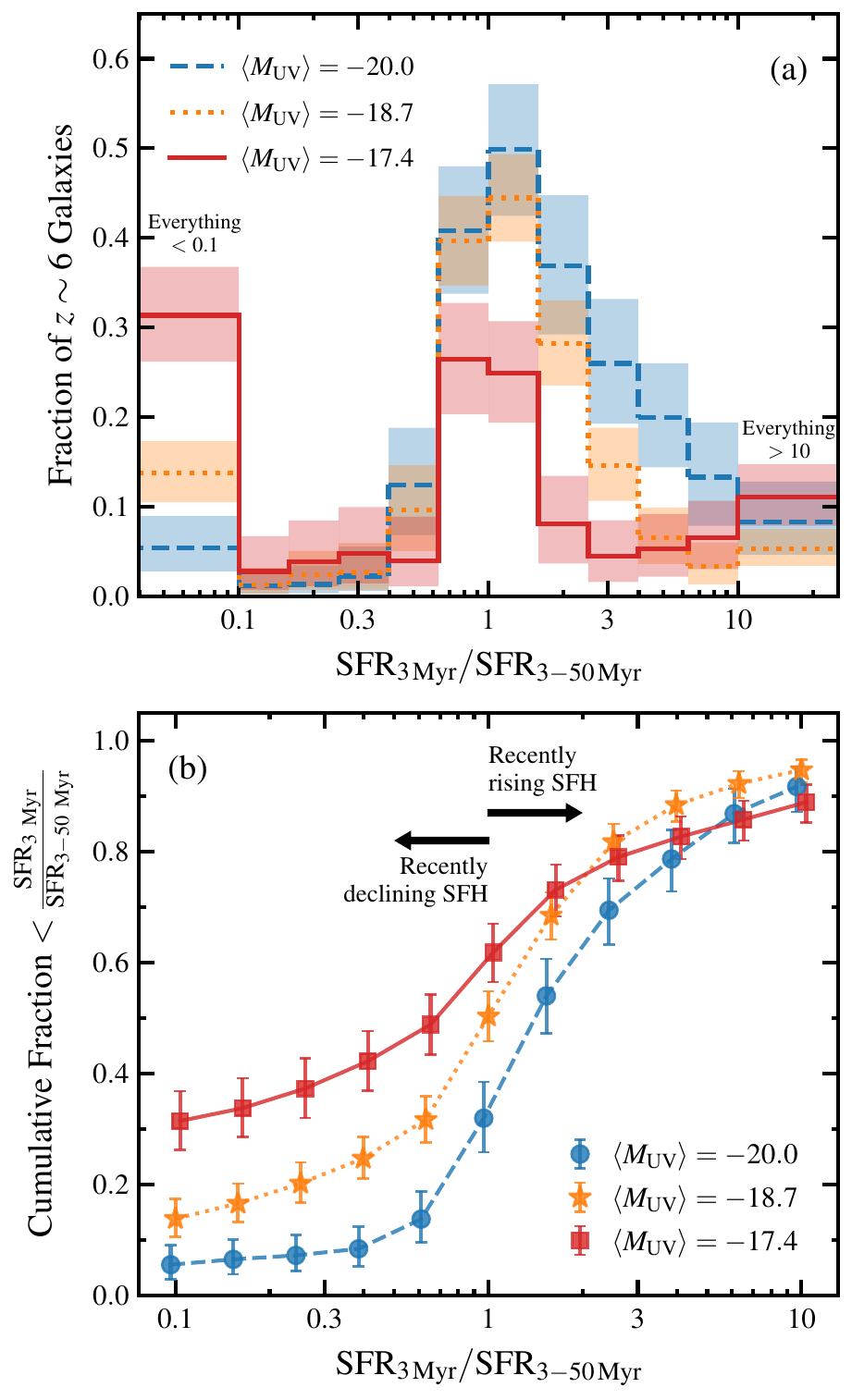}
\caption{The inferred probability (a) and cumulative (b) distribution functions of recent SFH shapes among $z\sim6$ LBGs as a function of UV luminosity. These plots encode more information than Fig. \ref{fig:dutyCycle}. In particular, they reveal that a considerably lower fraction of very UV-faint galaxies have had approximately constant SFRs over the past 50 Myr ($\SFRrat{} \approx 1$) compared to the more UV-luminous population. They also show that the shapes of the $\SFRrat{}$ distributions are considerably different for the UV-bright and very UV-faint bins. In panel (b), markers are slightly offset along the x-axis for clarity.}
\label{fig:SFRratPDFandCDF}
\end{figure}

These trends are qualitatively consistent with expectations of bursty SFHs.
The fraction of galaxies with strong recent SFR downturns should increase towards the faint end of the UV luminosity function (UVLF) because of how rapidly \Luv{} declines after a spike in SFR (see Fig. \ref{fig:SSPevolution}).
Within just 10 (30) Myr after a strong burst, \Luv{} will decrease by a factor of $\approx$10 ($\approx$25) if there is negligible subsequent star formation.
Therefore, the only UV-bright galaxies with strong recent SFR downturns will be exceptionally massive systems, and hence very rare given the exponential turnover in the stellar (and halo) mass function. 
Accordingly, the bright end of the $z\sim6$ UVLF should be dominated by galaxies with up-scattered \Luv{} (at fixed mass) due to a recent rise in SFR, consistent with our findings (Fig. \ref{fig:dutyCycle}).
Towards the faint end of the UVLF, galaxies are generally being sampled from a region of the mass function with a shallower slope.
This yields a more even mixture of relatively massive galaxies that have down-scattered in \Luv{} (due to an SFR downturn) and relatively low-mass galaxies that have up-scattered in \Luv{} (due to an SFR upturn) at the faint end. 

To provide more detailed information on how the distribution of recent SFH shapes changes with \Muv{}, we also compute the probability distribution function (PDF) and cumulative distribution function (CDF) in $\SFRrat{}$ (see Fig. \ref{fig:SFRratPDFandCDF}) using the same Bayesian formalism as above.
We find that a considerable fraction of $z\sim6$ LBGs have recent SFHs roughly consistent with constant SFR over the past 50 Myr (i.e., $\SFRrat{} \approx 1$).
This fraction falls significantly towards the faint end of our sample.
Among the UV-bright ($\langle \Muv{} \rangle = -20.0$) subset, $59\pm7$ percent of galaxies have $\SFRrat{}$ within $\pm0.2$ dex of unity.
This drops to $33^{+6}_{-5}$ percent in the very UV-faint ($\langle \Muv{} \rangle = -17.4$) subset.

The overall shapes of the $\SFRrat{}$ distributions also differ significantly at the bright and faint ends of our sample (see Fig. \ref{fig:SFRratPDFandCDF}).
Among the very UV-faint ($\langle \Muv{} \rangle = -17.4$) population, the majority of galaxies have either had roughly constant SFHs for the past 50 Myr, or a very strong downturn in SFR over the past 3 Myr.
UV-bright $z\sim6$ galaxies ($\langle \Muv{} \rangle = -20.0$) instead show a sharp cutoff towards even moderate SFR downturns, but have a long tail towards strong recent SFR upturns.

\section{Discussion} \label{sec:discussion}

The ubiquity of high EW \OIIIHb{} emission among UV-bright $z\sim7-8$ galaxies had long suggested that early galaxies were very often vigorously forming stars (sSFR$\sim$30 Gyr$^{-1}$; e.g., \citealt{Smit2014,Endsley2021_OIII,Stefanon2022_sSFR}).
It is now clear that these IRAC-based findings delivered a highly incomplete view of early star formation.
In this paper, we demonstrated that there is a substantial population of $z\sim6$ galaxies that have recently experienced a strong downturn in star formation rate ($\SFRrat{} < 0.2$).
These galaxies have low current sSFRs ($\lesssim$0.1--1 Gyr$^{-1}$; Figs. \ref{fig:veryLowsSFR10} and \ref{fig:GS80873}--\ref{fig:A2744_20941}), and are far more common at the faint end of the UVLF ($\Muv{} > -18$; see Fig. \ref{fig:dutyCycle}).
We also find that the fraction of $z\sim6$ galaxies experiencing a strong spike in SFR ($\SFRrat{} > 5$) remains fairly constant with UV luminosity over $-20 \lesssim \Muv{} \lesssim -17.5$ (Fig. \ref{fig:dutyCycle}).
In this section, we discuss what our results imply for the burstiness of star formation among early galaxies.
Along the way, we demonstrate that our results highlight an incompleteness issue faced by \textit{all} high-redshift galaxy samples, particularly among spectroscopic datasets, that must be overcome with future surveys to continue improving our understanding of galaxy formation. 

Because of how rapidly luminosity declines after a burst of star formation (Fig. \ref{fig:SSPevolution}), UV-faint galaxies are more representative of the diversity in recent SFH shapes among early galaxies.
These UV-faint systems include a more even mixture of relatively massive galaxies that have down-scattered in \Luv{} due to a recent decline in SFR, as well as relatively low-mass galaxies that have up-scattered in \Luv{} due to a recent spike in SFR (see \S\ref{sec:SFRratDistnsMuv}).
Accordingly, very UV-faint $z\sim6$ galaxies should span a wider range of overall galaxy properties (e.g., stellar mass, metallicity, size, ionizing photon production efficiency) compared to systems near the knee of the UVLF ($\Muv{} \sim -20$).
Put another way, it is less clear what the nature of a given $\Muv{} = -17.5$ $z\sim6$ galaxy is from its UV luminosity alone compared to a $z\sim6$ galaxy with $\Muv{} = -20$.

The fact that we see a significant fraction of very UV-faint $z\sim6$ galaxies at both extremes of the recent SFH shape distribution (see Fig. \ref{fig:SFRratPDFandCDF}) means that early galaxies frequently cycle through phases of rapid stellar mass assembly and other periods of much slower growth.
While in the rapid assembly phase, the sudden formation of many early-type O stars results in very efficient production of ionizing photons and, consequently, high-EW nebular line emission (e.g., \citealt{Withers2023,Atek2024,Boyett2024_jades}; \citetalias{Endsley2024}).
These starburst events appear to often be concentrated in compact ($<$20 pc) regions \citep[e.g.,][]{Vanzella2023,Adamo2024,Topping2024_RXCJ,Topping2024_A1703}, resulting in huge star formation rate surface densities where supernovae and radiative feedback efficiently drive out gas.
This process naturally results in subsequent periods of very low sSFR \citep[e.g.,][]{FaucherGiguere2018,Furlanetto2022}, which we are now seeing en masse with \JWST{} at the faint end (Fig. \ref{fig:dutyCycle}).

One major question moving forward is how long these periods of low sSFR typically last among $z\gtrsim6$ galaxies.
Not only is this timing sensitive to the feedback mechanisms regulating the growth of early galaxies, but it also has strong implications for understanding how the UVLF connects to the (stellar or halo) mass function.
Among our 23 LBGs with the most confident strong SFR downturns (\S\ref{sec:strongDownturnLBGs}), the majority are inferred to have experienced a strong burst of star formation $\approx$5--10 Myr ago (see Fig. \ref{fig:GS80873} for a spectroscopically-confirmed example).
But there is a tail to this distribution, where some of our LBGs seem to have spent the past $\approx$30--50 Myr with very low sSFR (see Fig. \ref{fig:veryLowsSFR10}).

These timescales imply that the burstiness of star formation at $z\sim6$ imposes a huge scatter in UV luminosity at fixed mass.
Galaxies with negligible star formation for the past $\approx$30--50 Myr (Fig. \ref{fig:veryLowsSFR10}) are likely $\approx$30$\times$ (i.e., $\approx$3--4 mags) fainter now in the far-UV continuum than during their most recent starburst phase (Fig. \ref{fig:SSPevolution}).
Even our LBGs that have only been relatively inactive for the past $\approx$5--10 Myr are likely $\approx$1--2 mags fainter now than during their most recent burst. 
This large scatter in \Luv{} helps explain the surprising abundance of relatively bright $z>10$ galaxies now being found with \JWST{} (e.g., \citealt{Bouwens2023,Harikane2023,Leung2023,PerezGonzalez2023,Casey2024,Donnan2024,Finkelstein2024,Robertson2024}).
Models have suggested that galaxies must scatter by $\approx$1--2 mags in \Muv{} at fixed halo mass to explain the measured $z>10$ UVLFs \citep[e.g.,][]{Mason2023,Mirocha2023,Munoz2023,Shen2023,Sun2023_brightEnd,Kravtsov2024}.
Our findings demonstrate that such fluctuations in \Muv{} are fairly common even at $z\sim6$, where we can currently obtain the best statistical constraints on SFR stochasticity at high redshifts by using NIRCam to probe the Balmer break amplitudes, the \OIII{} EWs, and the H$\alpha$ EWs.
Whether the burstiness at $z\sim6$ implied by our findings can precisely explain the $z>10$ UVLF measurements is beyond the scope of this work, but it is clear that huge variations in \Luv{} from rapid ($\sim$3--30 Myr) strong SFR upturns and downturns are consistent with our results.

In our $z\sim6$ LBG sample, short ($\approx$5--10 Myr) periods of low sSFR are more common than relatively long periods ($\gtrsim$30 Myr).
At face value, this is consistent with a scenario wherein $z\sim6$ galaxies typically cycle through bursts on a very rapid ($\sim$5--10 Myr) cadence, as would be expected if strong feedback acts almost immediately after the onset of star formation (via, e.g., radiative winds).
However, this implied timing of bursts could very well be due to an incompleteness issue faced by not only our LBG sample, but every empirical high-redshift galaxy sample.

After a burst of star formation, galaxies rapidly become much fainter in both continuum and line emission as the most massive stars (with the highest light-to-mass ratios) quickly die off (Fig. \ref{fig:SSPevolution}).
For this reason, galaxies with longer recent stretches of low sSFR are much more challenging to detect (and more-so characterize) at fixed mass.
This naturally explains why every one of our LBGs with very low sSFR for the past $\geq$10 Myr (Fig. \ref{fig:veryLowsSFR10}) is either relatively massive ($\Mstar{} \sim 10^9\,\Msol{}$) or very highly magnified ($\mu > 10$).
The same is true of the three similar spectroscopically-confirmed $z>5$ systems reported in the literature \citep{Strait2023,Looser2024,Weibel2024}.
Without very strong lensing, existing data is simply not sensitive to low-to-moderate mass ($\Mstar{} \lesssim 10^8\,\Msol{}$) $z\gtrsim6$ galaxies that have been relatively inactive for tens of Myr (see Fig. \ref{fig:SSPevolution}).
Objects with ongoing bursts are far easier to detect (and characterize) at low masses because the prevalence of early-type O stars makes them very bright in continuum and line emission.

A notable outcome of this effect is that different high-redshift galaxy selection techniques will preferentially identify systems with different recent SFH shapes, and hence come to different conclusions on the typical characteristics of early galaxies.
Spectroscopic samples selected on nebular line detections, and even photometric redshift samples folding in long-wavelength NIRCam data (with unique color patterns created by line emission) are strongly weighted towards galaxies in the rapid growth phase.
These samples will give the impression that galaxies at fixed mass are more efficient at producing ionizing photons than is true if considering the full distribution in recent SFH shapes.
LBG selections like those used in this work are less biased against objects with relatively long periods of low sSFR since the stellar continuum fades less rapidly than nebular emission after a burst of star formation.
Nonetheless, every flux-limited high-redshift sample will suffer from more incompleteness to galaxies with longer recent periods of low sSFR.

For this reason, we do not yet have a clear, comprehensive picture of the duty cycle of star formation in even moderately-massive ($\Mstar{} \sim 10^8\,\Msol{}$) $z\sim6$ galaxies.
Indeed, \citet{Sun2023_observability} predicted that even the deep JADES NIRCam imaging is $<$50\% complete to detecting $\Mstar{} = 10^{8-8.5}\,\Msol{}$ galaxies at $z=6-8$ using the second-generation Feedback in Realistic Environments (\textsc{fire}-2; \citealt{Hopkins2018_FIRE2}) simulations \citep{Ma2018_burstySFH,Ma2019,Ma2020} which have bursty SFHs.
The selection completeness of these moderately-massive systems (either via Lyman-break, photo-$z$, or spectroscopic line emission techniques) will of course be even lower than the detection completeness.

But these persisting empirical challenges do not diminish the tremendous progress that has been made with the first two years of \JWST{} data.
It is now clear that early galaxies do often undergo dramatic (factor of $\gtrsim$5) upturns \textit{and downturns} in their SFR on fairly short ($\sim$10 Myr) timescales, with some having formed very little stellar mass for the past $\sim$30--50 Myr (\S\ref{sec:strongDownturnLBGs}; \S\ref{sec:specDownturns}; \citealt{Strait2023,Looser2024,Weibel2024}).
This is leagues beyond where our empirical understanding stood with IRAC.
We note that there are also efforts to identify early galaxies that had low sSFR up until very recently when they `rejuvenated' \citep{Witten2024}.
While this is a promising approach toward better understanding the duty cycle of early star formation, this is challenging because the new O stars will easily outshine the older population, and hence diminish their empirical signatures.
But this outshining effect can be alleviated with spatially-resolved SFH constraints on individual clumps within early galaxies, particularly with the assistance of gravitational lensing \citep{GimenezArteaga2023,GimenezArteaga2024}.

In the future, extremely deep, pencil-beam surveys will be required to better understand how long low-to-moderately massive ($\Mstar{}\lesssim10^{7-8}\,\Msol{}$) $z\gtrsim6$ galaxies spend in phases of low sSFR.
The Cycle 2 GO program GLIMPSE (PID 3293; PIs Hakim \& Chisholm) will take a key next step towards this goal by delivering extremely deep multi-band NIRCam imaging (including in F410M and F480M) in the Abell S1063 lensing field, pushing to intrinsic magnitudes of $m\gtrsim32$.
Combined with wider-area studies sensitive to the fairly massive ($\Mstar{} \sim 10^{9-10}\,\Msol{}$) population, the community can build a much more complete census of recent SFHs among early galaxies across multiple orders of magnitude in mass. 
This will greatly advance our understanding of the physical processes governing galaxy formation, as well as help resolve why $z\gtrsim12$ galaxies are surprisingly abundant.

\section{Summary} \label{sec:summary}

We use very deep broad+medium-band NIRCam imaging in the GOODS and Abell 2744 fields to investigate the recent SFHs of a large ($N = 368$) sample of $z\sim6$ Lyman-break galaxies spanning (de-lensed) absolute UV magnitudes of $-22 \lesssim \Muv{} \lesssim -16$.
We focus on $z\sim6$ as this is the highest redshift where we can obtain statistical constraints on key SFH diagnostics encoded in the SED (the Balmer break amplitude, \OIII{} EW, and H$\alpha$ EW) with NIRCam imaging.
Our main conclusions are as follows.

\begin{enumerate}

    \item From the posteriors of \textsc{beagle} SED fits, we systematically identify 23 $z\sim6$ LBGs that have confidently ($>$90\% probability) experienced a strong recent downturn in star formation activity ($\SFRrat{} < 0.2$; see \S\ref{sec:strongDownturnLBGs} and Table \ref{tab:properties}). Four of these galaxies show evidence of very low sSFRs ($\lesssim$0.2 Gyr$^{-1}$) for the past $\approx$10--50 Myr (see Fig. \ref{fig:veryLowsSFR10}), similar to the `mini-quenched' systems that have been confirmed with spectra \citep{Strait2023,Looser2024,Weibel2024}. Several other LBGs in our sample convincingly show strong Balmer breaks ($\BBrat{} > 1.4$) and very weak lines (\OIIIHb{} and H$\alpha$ EWs$\lesssim$100 \AA{}), or weak Balmer breaks and moderate line EWs ($\sim$200--400 \AA{}).
    
    \item Using public NIRSpec/prism data for 25 of our LBGs, we verify that our photometric \textsc{beagle} SED fits robustly recover the redshifts, Balmer break amplitudes, \OIIIHb{} EWs, and H$\alpha$ EWs (see Fig. \ref{fig:specComparison}). For this reason, we find good agreement between the recent SFH shapes inferred from photometry alone and those inferred using the combined spectro-photometric constraints (Fig. \ref{fig:specComparison_SFRrat}). Three of the 23 candidates with the most confident recent strong SFR downturns have NIRSpec spectra, and none show clear signs of extremely efficient ionizing photon escape or differential dust extinction that might explain weak nebular lines with high sSFR. The spectroscopic measurements of these three galaxies firmly support the interpretation that each experienced a strong downturn in SFR $\approx$5--30 Myr ago (see Figs. \ref{fig:GS80873}--\ref{fig:A2744_20941}).

    \item Across our full LBG sample, we infer H$\alpha$ to far-UV luminosity ratios spanning at least 2 orders of magnitude (see Fig. \ref{fig:Lrat_Muv}a). Nineteen LBGs show very low luminosity ratios of $\logLrat{} \leq  -2.5$ (including three systems with NIRSpec/prism measurements), similar to several local dwarf galaxies that have bursty SFHs (e.g., \citealt{Emami2019}). The median \Lrat{} ratio of our sample decreases significantly with UV luminosity (Fig. \ref{fig:Lrat_Muv}b), the opposite trend as expected from metallicity effects alone. This decline in \Lrat{} with \Luv{} is instead expected from bursty SFHs because both the UV continuum luminosity and (to a stronger degree) the H$\alpha$ luminosity will rapidly weaken as the most massive stars die off following a burst. We also find that very low \Lrat{} ratios are 5$\times$ more common among the faintest subset of our LBG sample ($\langle \Muv{}\rangle = -17.4$) compared to the brightest subset ($\langle \Muv{}\rangle = -20.0$; see Fig. \ref{fig:Lrat_Muv}b), again qualitatively consistent with expectations of bursty SFHs.

    \item From the SFH posteriors output by our \textsc{beagle} SED fits, we find that strong recent SFR downturns ($\SFRrat{} < 0.2$) are far more common among the faintest galaxies in our sample (see Fig. \ref{fig:dutyCycle}). In the brightest subset ($\langle \Muv{}\rangle = -20.0$), this fraction is only $\approx$7\%, but rises by a factor of $\approx$5 in the faintest subset ($\langle \Muv{}\rangle = -17.4$). We find no significant change in the frequency of strong recent SFR upturns ($\SFRrat{} > 5$) with UV luminosity (see Fig. \ref{fig:dutyCycle}). These trends are qualitatively consistent with expectations of bursty SFHs given 1) the shape of the mass function and 2) how rapidly \Luv{} declines after a burst of star formation (see the discussion in \S\ref{sec:SFRratDistnsMuv}).

    \item Our results indicate a far greater diversity in the distribution of recent SFH shapes among very UV-faint $z\gtrsim6$ galaxies compared to the most UV-luminous systems (see Fig. \ref{fig:SFRratPDFandCDF}). This in turn implies that very UV-faint galaxies at high-redshift span a wider range of overall galaxy properties (e.g., stellar mass, metallicity, size, ionizing photon production efficiency) compared to systems near the knee of the UVLF. The fact that we see a significant fraction of very UV-faint $z\sim6$ galaxies at both extremes of the recent SFH distribution means that early galaxies frequently cycle through phases of rapid stellar mass assembly and other periods of much slower growth.

    \item Our LBG sample includes several objects with low sSFR for the past $\approx$5--50 Myr (e.g., Figs. \ref{fig:veryLowsSFR10} and \ref{fig:GS80873}--\ref{fig:A2744_20941}). This implies that the burstiness of star formation at $z\sim6$ results in huge ($\gtrsim$1--2 mag) scatter in \Muv{} at fixed mass (see Fig. \ref{fig:SSPevolution}). This scatter helps explain the surprising abundance of $z>10$ galaxies now being found with \JWST{}. 

    \item Short ($\approx$5--10 Myr) periods of low sSFR appear more common than relatively long periods ($\gtrsim$30 Myr). However, this may very well be due to an incompleteness issue faced by not only our LBG sample, but every empirical high-redshift galaxy sample. Because a galaxy's luminosity declines so rapidly after a burst of star formation (Fig. \ref{fig:SSPevolution}), it is far more challenging to detect (and more-so identify and characterize) galaxies with longer recent stretches of low sSFR. Without very strong lensing ($\mu \gtrsim 10$), existing data is simply not sensitive to low-to-moderate mass ($\Mstar{} \lesssim 10^8\,\Msol{}$) $z\sim6$ galaxies that have been relatively inactive for tens of Myrs (see also \citealt{Sun2023_observability}). Consequently, high-redshift samples selected on line emission (e.g., spectroscopic samples or photo-$z$ selected samples folding in long-wavelength NIRCam photometry) will be strongly weighted towards galaxies in the rapid growth phase with high ionizing photon production efficiency. Extremely deep data is required to build a more comprehensive understanding of early galaxy assembly, as well as the impact of low-mass versus high-mass galaxies to cosmic reionization.
    
\end{enumerate}

%% IMPORTANT! The old "\acknowledgment" command has be depreciated. It was
%% not robust enough to handle our new dual anonymous review requirements and
%% thus been replaced with the acknowledgment environment. If you try to 
%% compile with \acknowledgment you will get an error print to the screen
%% and in the compiled pdf.
%% 
%% Also note that the akcnowlodgment environment does not support long amounts of text. If you have a lot of people and institutions to acknowledge, do not use this command. Instead, create a new \section{Acknowledgments}.
\section{Acknowledgments}
RE thanks Claude-Andr{\'e} Faucher-Gigu{\`e}re, Julian Mu{\~n}oz, Harley Katz, Gourav Khullar, Jenna Samuel, and Jason Sun for enlightening conversations that improved this work. 
RE also thanks Lukas Furtak for help with the lensing interpretation for A2744-23813 and A2744-23814.
This research is based on observations made with the NASA/ESA Hubble Space Telescope obtained from the Space Telescope Science Institute, which is operated by the Association of Universities for Research in Astronomy, Inc., under NASA contract NAS 5–26555.
This work is based [in part] on observations made with the NASA/ESA/CSA James Webb Space Telescope. The data were obtained from the Mikulski Archive for Space Telescopes at the Space Telescope Science Institute, which is operated by the Association of Universities for Research in Astronomy, Inc., under NASA contract NAS 5-03127 for JWST.
We thank the FRESCO, JOF, GLASS, UNCOVER, MegaScience, ALT, and MAGNIF teams, along with the teams lead by (co-)PIs W. Chen, T. Treu, \& H. Williams (PID 2756), E. Iani (PID 3538), and T. Morishita (PID 3990) for developing their observing program with a zero-exclusive-access period.
We are grateful for the collective contributions of the approximately 20,000 humans around the world who designed, built, tested, commissioned, and operate JWST.
Some of the data products used herein were retrieved from the Dawn JWST Archive (DJA). DJA is an initiative of the Cosmic Dawn Center (DAWN), which is funded by the Danish National Research Foundation under grant DNRF140.
The authors acknowledge the Texas Advanced Computing Center (TACC) at The University of Texas at Austin for providing HPC resources that have contributed to the research results reported within this paper.
This material is based in part upon High Performance Computing (HPC) resources supported by the University of Arizona TRIF, UITS, and Research, Innovation, and Impact (RII) and maintained by the UArizona Research Technologies department.

%% To help institutions obtain information on the effectiveness of their 
%% telescopes the AAS Journals has created a group of keywords for telescope 
%% facilities.
%
%% Following the acknowledgments section, use the following syntax and the
%% \facility{} or \facilities{} macros to list the keywords of facilities used 
%% in the research for the paper.  Each keyword is check against the master 
%% list during copy editing.  Individual instruments can be provided in 
%% parentheses, after the keyword, but they are not verified.

\vspace{5mm}
\facilities{HST(ACS), JWST(NIRCam and NIRSpec)}

%% Similar to \facility{}, there is the optional \software command to allow 
%% authors a place to specify which programs were used during the creation of 
%% the manuscript. Authors should list each code and include either a
%% citation or url to the code inside ()s when available.

\software{\textsc{numpy} \citep{harris2020_numpy},
\textsc{matplotlib} \citep{Hunter2007_matplotlib},
\textsc{scipy} \citep{Virtanen2020_SciPy},
\textsc{astropy} \citep{astropy:2013, astropy:2018},
\textsc{jwst} \citep{Bushouse2023},
\textsc{Source Extractor} \citep{Bertin1996},
\textsc{sep} \citep{Barbary2016_sep},
\textsc{photutils} \citep{Bradley2022_photutils},
\textsc{beagle} \citep{Chevallard2016},
\textsc{multinest} \citep{Feroz2008,Feroz2009},
\textsc{emcee} \citep{ForemanMackey2013_emcee},
\textsc{bpass} \citep{Stanway2018},
\textsc{cloudy} \citep{Ferland2013},
\textsc{grizli} \citep{grizli2023},
\textsc{parsec} \citep{Bressan2012_parsec,Chen2015_parsec}
}

%% Appendix material should be preceded with a single \appendix command.
%% There should be a \section command for each appendix. Mark appendix
%% subsections with the same markup you use in the main body of the paper.

%% Each Appendix (indicated with \section) will be lettered A, B, C, etc.
%% The equation counter will reset when it encounters the \appendix
%% command and will number appendix equations (A1), (A2), etc. The
%% Figure and Table counter will not reset.

\appendix

\section{Potential Influence of Binary Stellar Evolution} \label{app:binaries}

\begin{figure*}
\includegraphics[width=\textwidth]{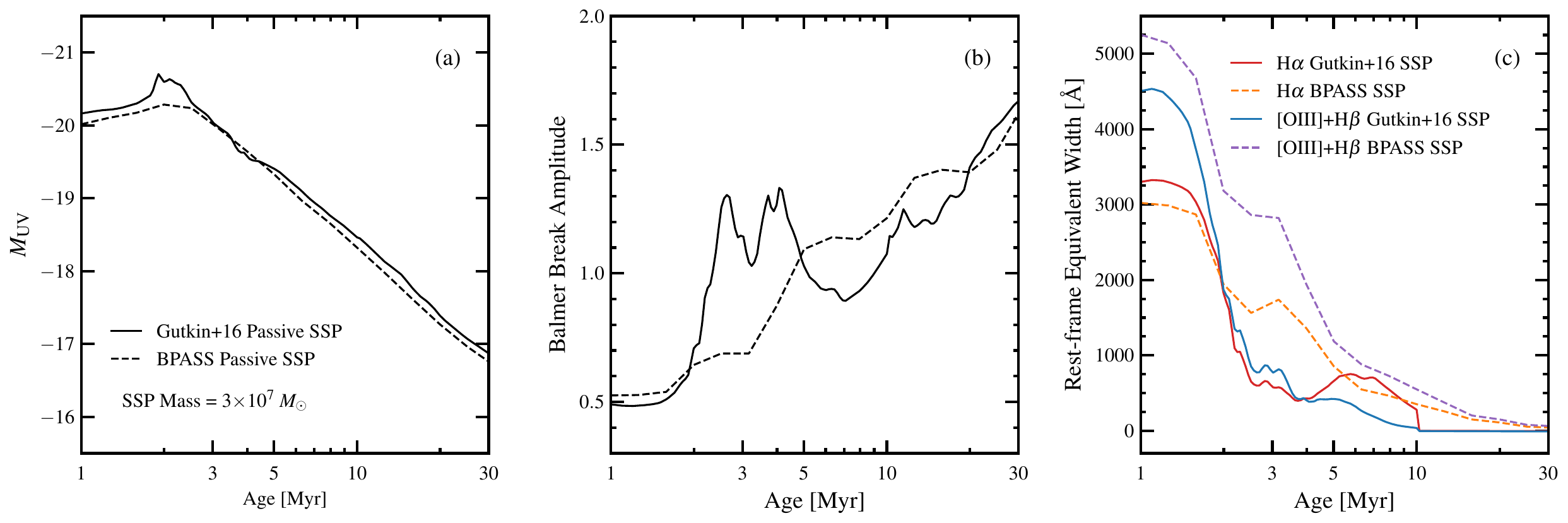}
\caption{Comparison of time evolution in SED properties between the single-star \citet{Gutkin2016} photoionization models implemented in \textsc{beagle} with binary evolution \textsc{bpass} photoionization models. All models adopt log $U = -2.5$, $Z \approx 0.1\ Z_\odot$, and no dust attenuation. The binary evolution prescription in \textsc{bpass} extends the lifetime of very high EW ($>$1000 \AA{}) nebular line emission to $\approx$5 Myr after an SSP is formed, and allows for significant (EW$\approx$100 \AA{}) line emission for up to $\approx$20 Myr.}
\label{fig:BPASS}
\end{figure*}

Binary stellar evolution can significantly extend the main-sequence lifetimes of massive stars via mass transfer from the secondary star \citep[e.g.,][]{Sana2012}.
Consequently, accounting for binary stellar evolution will prolong the timescale over which nebular emission remains strong after a burst of star formation.
We quantify the potential scale of this effect using the Binary Population and Spectral Synthesis (\textsc{bpass}) v2.2.1 models that have been run through \textsc{cloudy} to produce SEDs that include nebular emission.\footnote{\url{https://bpass.auckland.ac.nz/4.html}}

Fig. \ref{fig:BPASS} compares the time evolution in far-UV luminosity, Balmer break amplitude, as well as H$\alpha$ and \OIIIHb{} EWs predicted by the \textsc{bpass} models with the single-star \citet{Gutkin2016} photoionization models implemented in \textsc{beagle}. 
As expected, the extended lifetime of early-type O stars in the \textsc{bpass} models prolongs the timescale for very high-EW ($>$1000 \AA{}) H$\alpha$ and \OIIIHb{} emission, lasting up to $\approx$5 Myr after a strong burst of star formation as opposed to $\approx$2 Myr with the single-star \citet{Gutkin2016} models (Fig. \ref{fig:BPASS}c).
Moreover, significant (EW$\approx$100 \AA{}) \OIIIHb{} and H$\alpha$ emission persists for 20 Myr after the formation of an SSP with \textsc{bpass}, while all nebular emission ceases to exist after 10 Myr in the \citet{Gutkin2016} models.

The prolonged lifetime of massive stars in the \textsc{bpass} models would concievably also keep the UV luminosity higher for longer, as well as weaken the Balmer break amplitude at fixed SSP age. 
However, we find that there are very little differences in the evolution in the far-UV luminosity between the two models (Fig. \ref{fig:BPASS}a), as well as relatively little differences in the evolution of the Balmer break amplitude (Fig. \ref{fig:BPASS}b).
The \citet{Gutkin2016} models show a more prominent bump in UV luminosity at $\approx$2 Myr after the formation of an SSP, and moreover shows jumps in the Balmer break amplitude at $\approx$3--5 Myr that are not seen with the \textsc{bpass} models.
These differences are likely due, at least in part, to different prescriptions for the post-main-sequence evolution of massive stars (S. Charlot, priv. comm.).
A detailed comparison between model prescriptions is beyond the scope of this work (see, e.g., \citealt{Plat2019,Lecroq2024}) as we simply aim to illustrate the potential qualitative impact of binary stellar evolution on our conclusions.

Qualitatively, accounting for binary stellar evolution would require less rapid upturns in SFR for the extreme emission line population. 
However, it would require even longer timescales of low sSFR for the subset of galaxies with surprisingly weak nebular line EWs (\S\ref{sec:strongDownturnLBGs}, \S\ref{sec:specDownturns}, \citealt{Strait2023,Looser2024,Weibel2024}).
A quantitative investigation of the potential impact of binary stellar evolution on the recent SFHs of $z\sim6$ LBGs is left for future work.

\bibliography{main}{}
\bibliographystyle{aasjournal}

%% This command is needed to show the entire author+affiliation list when
%% the collaboration and author truncation commands are used.  It has to
%% go at the end of the manuscript.
%\allauthors

%% Include this line if you are using the \added, \replaced, \deleted
%% commands to see a summary list of all changes at the end of the article.
%\listofchanges

\end{document}